\documentstyle[aps,prb,preprint]{revtex}
\begin{document}

\title{NMR and NQR Fluctuation Effects in Layered
Superconductors}
\author{D. Fay$^1$ , J. Appel$^1$, C. Timm$^2$, 
and A. Zabel$^1$}
\address{$^1$I. Institut f\"ur Theoretische Physik,
Universit\"at Hamburg, Jungiusstr. 9, 20355 Hamburg, 
Germany}
\address{$^2$Institut f\"ur Theoretische Physik, 
Freie Universit\"at Berlin,Arnimallee 14,
14195 Berlin,  Germany}
\date{\today}
\maketitle
\begin{abstract}
We study the effect of thermal fluctuations of the s-wave order
parameter of a quasi two dimensional superconductor on the nuclear
spin relaxation rate near the transition temperature $T_C$. We consider
both the effects of the amplitude fluctuations and the 
Berezinskii-Kosterlitz-Thouless (BKT) phase fluctuations in weakly coupled 
layered superconductors. In the treatment of the amplitude fluctuations we 
employ the Gaussian approximation and evaluate the longitudinal relaxation 
rate $T_1^{-1}$ for a clean s-wave superconductor, with and without pair 
breaking effects, using the static pair fluctuation propagator $\cal{D}$.  
The increase in $T_1^{-1}$ due to pair breaking in $\cal{D}$ is 
overcompensated by the decrease arising from the single particle Green's 
functions. The result is a strong  effect on $T_1^{-1}$ for even a small amount 
of pair breaking. The phase fluctuations are described in terms of dynamical 
BKT excitations in the form of 
pancake vortex-antivortex (VA) pairs. We calculate the effect of the magnetic 
field fluctuations caused by the translational motion of VA excitations on 
$T_1^{-1}$ and on the transverse relaxation rate $T_2^{-1}$ on both sides of 
the BKT transition temperature $T_{\text{BKT}}<T_C$. The results for the NQR 
relaxation rates depend strongly on the diffusion constant $D$ that governs 
the motion of free and bound vortices as well as the annihilation of VA pairs. 
We discuss the relaxation rates for real multilayer systems where 
D can be small and thus increase the lifetime of a VA pair, leading to an 
enhancement of the rates. We also discuss in some detail the experimental 
feasibility of observing the effects of amplitude fluctuations in layered s-wave
superconductors such a the dichalcogenides and the effects of phase 
fluctuations in s- or d-wave superconductors such as the layered cuprates.
\end{abstract}
\vspace{0.5in}
\section{INTRODUCTION}

	The most common NMR and NQR experiments on high-$T_C$ and other 
quasi two dimensional superconductors concern the Knight shift and the 
longitudinal and transverse relaxation rates.  Both of these experiments 
explore the low-frequency spin dynamics of electrons and holes in normal 
metals and superconductors.  The relaxation rates are caused by 
the time dependence of the fluctuating magnetic fields.  In superconductors 
these field fluctuations originate from electronic quasiparticle excitations and 
from the motion of magnetic vortices.  In the high-$T_C$ cuprates the 
quadrupolar Cu spin-lattice relaxation can be due to the transitions between 
quadrupolar states of the Cu nuclei caused by the interaction of the nuclear 
spins with the time dependent magnetic fields of the vortices.  The 2D cuprates 
such as Bi-2212 consist of CuO$_2$ layers with a very small interlayer hopping 
matrix elememt  $t_\perp$.  \cite{Slichter}  The magnetic field fluctuations near 
the real superconducting transition temperature $T_{C}\,$, where the resistivity 
goes to zero and the long-range order is established by Josephson phase 
coupling,  are caused by both the quasiparticle excitations of the normal and 
superconducting states and by the spontaneous excitation of thermal 
vortex-antivortex (VA) pairs.  These vortex excitations occur close to the 
Berezinskii-Kosterlitz-Thouless (BKT) transition temperature of a 2D layer,  
$T_{\text{BKT}} < T_C$ ,  cf. Fig. 1.  \cite{Halperin,Blatter,Glaz}  Whereas the 
long-wavelength Gaussian fluctuations of the quasiparticle excitations consist 
of amplitude and long-wavelength phase fluctuations of the complex order 
parameter,  the VA fluctuations are primarily phase fluctuations of the order 
parameter.  The BKT scenario also applies to 2d layers that are weakly 
coupled by electromagnetic and Josephson coupling effects. 
\cite{CTlay,Friesen,Corson}  
Around the real transition temperature $T_C$ there exists the narrow Ginzburg 
regime of critical fluctuations where the Josephson and Coulomb couplings 
into the third dimension begin to establish the phase coherence between 
neighboring layers. This is representated by the shaded area around $T_C$ in 
Fig. 1.  According to the detailed calculations of the specific heat fluctuations of 
Ramallo and Vidal \cite{Ramallo} and their comparison with experimental 
results,  the three-dimensional critical behavior is restricted to a rather narrow 
range in the cuprates, $t\equiv |T-T_{C}| < 10^{-2}$,  so that a wide region of 
Gaussian fluctuations of the order of 10K exists above $T_C$.  The same 
considerations should be be valid for the dichalcogenides which are layered 
systems exhibiting s-wave pairing. The BKT vortex-antivortex fluctuations 
exist above $T_C$ below the mean field transition temperature 
$T_{C0}$ (cf. Fig. 1).  At  $T_{cr} > T_{C}$, the interlayer phase coherence 
becomes so weak that the Josephson vortices proliferate and the 3D phase 
coupling ceases to exist.

In Sec. II we study the effect of Gaussian fluctuations on the spin-lattice 
relaxation rate  of s-wave superconductors near the transition temperature for 
layered systems.  The NMR spin relaxation rate  T$_1^{-1}$  has been 
investigated intensively in the high-$T_C$ superconductors.  At present however 
the general opinion seems to be that only the electron-doped high-$T_C$ 
superconductors may exhibit s-wave pairing. \cite{Tsuei,Manskepre} Other 
layered compounds with s-wave pairing are the layered transition metal 
dichalcogenides. Long wavelength thermal fluctuations of the superconducting 
order parameter,  i.e., Gaussian fluctuations, are expected to play a pertinent 
role in layered compounds for several reasons;  in particular,  the quasi 2D 
structure of these metallic systems, the small coherence length $\xi_0$ 
in the layers (cuprates),  and the large penetration depth $\lambda$ for fields 
parallel to the layers.  The effect of such fluctuations on $T_1^{-1}$ in these 
and similar systems has been the subject of numerous  papers. 
\cite{Wegeretal,ManivAlex,Kuboki,Heym,Kautz,AppelFay,RandVarl,%
VarlReview,Carreta1,Carreta2,Eschrig}  
Most of the previous authors 
\cite{Kuboki,Heym,RandVarl,VarlReview,Carreta1,Eschrig} included 
non-magnetic impurity scattering and also set the small external NMR 
frequency $\omega$  equal to zero. In such theories it is not clear whether the 
correct clean limit is obtained.  For example,  in a clean system without pair 
breaking,  we find $1/T_1\propto 1/\omega\,$,  for $T\rightarrow T_C$ .  To 
elucidate this limit we consider a clean system from the outset.  This may be a 
reasonable first approximation for systems like the cuprates where $\xi_0$  is 
large compared with the mean free path $\ell$.

Certainly for the cuprates,  and many other layered systems as well,  it is 
important to include the pair breaking effect of inelastic scattering due to 
exchange of low energy Bosons,  spin fluctuations or phonons,  for example.  
To do this correctly would require solution of the strong-coupling Eliashberg 
equations taking full account of the energy dependence of the pairing 
interaction and the single-particle self energy.  To date no such theory has 
appeared in the literature with respect to NMR in layered systems for s-wave 
pairing. \cite{Tewordt}  In the previous weak-coupling theories inelastic 
scattering has usually been accounted for by introducing a cutoff in the pair 
fluctuation propagator or a constant lifetime in the single particle Green's 
function.  We also do a weak-coupling calculation and simulate the effect of 
inelastic scattering through a pair breaking parameter in analogy to magnetic 
impurity scattering.  The large qualitative and quantitative effect of pair breaking 
indicates that the results of a weak coupling calculation should be viewed with 
caution until they can be confirmed with a strong coupling Eliashberg 
calculation.  Since the inclusion of pair breaking in the clean case qualitatively 
changes the results we give some of the details both with and without pair 
breaking.

In Sec. III  we study the effect of topological intra-layer phase fluctuations 
(vortices) on the NQR spin-lattice and spin-spin relaxation rates $T_{1}^{-1}$ 
and $T_{2}^{-1}$  .  In contrast to the Gaussian fluctuations of Sec. II,  which 
contain the long-wavelength fluctuations of the amplitude of the order 
parameter,  the topological fluctuations are,  primarily,  the 2D 
singular phase fluctuations of the superconducting order parameter. 
Experimentally,  the relaxation rates $T_{1,2}^{-1}$ in the cuprates can be 
obtained in zero magnetic field from NQR spin-echo experiments on  
$^{63}$Cu and $^{65}$Cu nuclei. We consider NQR to allow direct application 
of the zero-field BKT theory.  The translational motion of vortices and 
antivortices and the corresponding time dependent magnetic fields not only 
affect the nuclear spin relaxation but also the NQR line width and for this 
reason we calculate,  in addition to $T_{1}^{-1}$ , the spin-phase relaxation rate 
\mbox{$T_{2}^{-1}$ .}  The distribution of VA pairs is determined by the singular 
parts of the otherwise smooth phase fields of the complex order parameter.  
The topological VA excitations reduce the transition temperature of a single 
layer to the Berezinskii-Kosterlitz-Thouless temperature $T_{\text{BKT}}$ ,  
above which the bound VA pairs begin to break up into free pancake vortices 
and antivortices.  We study, in the temperature regime above 
$T_{\text{BKT}}$, the magnetic field correlation functions that determine 
$T_{1}^{-1}$  and $T_{2}^{-1}$ .  The magnetic field fluctuations are caused by 
the translational motion of the thermally excited vortices and antivortices.  
The diffusion constant of the vortices and antivortices,  $D(T)$,  plays a crucial 
role for the magnitudes of the relaxation rates;  it determines both the free vortex 
and antivortex motions and the recombination time of a VA pair.  We find that $D$ 
must be sufficiently small in order to get vortex relaxation rates $T_{1,2}^{-1}$  
that are comparable to those caused by quasiparticle relaxation.  The latter 
cause relaxation rates of the order of  $10^{2}$ to $10^{3}$ $s^{-1}$ near 
$T_C$ of the cuprates.  The effect of the time-varying magnetic fields due to the 
VA fluctuations on $T_{2}^{-1}$  is also of some interest for the following 
reason:  Of the two mechanisms causing the transverse relaxation in the 
cuprates,  namely the indirect nuclear spin-spin coupling and the spin-lattice 
relaxation, we expect that only the second contribution is affected by the VA 
fluctuations.  These fluctuations lead to an exponential decay of the time 
dependent transverse magnetization of the $^{63}$Cu or $^{65}$Cu nuclear 
spins. \cite{Penn}  On the other hand,  the individual $^{63}$Cu - $^{63}$Cu  
indirect coupling leads to a Gaussian time decay and is of such short range, 
$r \ll \xi_{0}$ ,  \cite{Gorney}  that this coupling remains alltogether unaffected 
by the transition into the superconducting state.

In Sec. IV we discuss the experimental situation in some detail with respect to 
the possibility of observing the effect of superconducting fluctuations on the 
longitudinal and transverse relaxation rates of the nuclear spins in layered 
systems such as the cuprates with s-wave pairing and conventional layered 
superconductors such as the transition metal dichalcogenides.%
\vspace{0.2in}

\section{ EFFECT\ OF\ PAIRING\ FLUCTUATIONS\ ON\ $T_1^{-1}$}

\subsection{General Formalism}

The NMR spin relaxation rate is given by
\begin{equation}
\frac{1}{T_1T}=\sum_{\bf q} A_{\bf q} \frac{\mbox{Im}\chi(\bf q,\omega)}
{\omega} \,,
\label{T1Def}
\end{equation}
where $\chi$  is the transverse (spin-flip) susceptibility,  $\omega$  is the 
external frequency,  and $A_{\bf q}$ is determined by the fine structure 
constants and should be large for the q's that couple to large values of 
$\mbox{Im}\chi(\bf q,\omega)$. Here we assume that $A_{\bf q}$  is a constant.
The susceptibility diagrams through first order in the Cooper pair fluctuation 
propagator,  $\cal{D}$,  are shown in Fig. 2. The zeroth order contribution,  
$\chi^{(0)}$   yields Korringa's law in the normal phase and,  in the absence of 
pair-breaking,  the Hebel-Slichter peak below $T_C$ for s-wave pairing. The 
leading fluctuation contributions are given by the diagrams $\chi_{FF}$   and 
$\chi_{GG}$ . We follow the notation of Maniv and Alexander \cite{ManivAlex} ,
who considered the clean case in 3D,  and write 
\begin{equation}
\sum_{\bf q}\mbox{Im}\chi({\bf q},\omega) = \mbox{Im}\chi(\omega)  , 
\label{Def2}
\end{equation} 
where
\begin{equation}
\mbox{Im}\chi(\omega) = \mbox{Im}\chi_{FF}(\omega) +
\mbox{Im}\chi_{GG}(\omega) \, .
\end{equation}
The first term represents the Maki-Thompson (MT) diagram, 
\cite{Maki,Thompson}  $\chi_{FF}=\chi_{MT}$,  which also has a long history in 
the calculation of the fluctuation conductivity.  The second term accounts for the 
self energy, or density of states (DOS), effect that is caused by the 
renormalization of the normal state Green's functions by superconducting 
fluctuations.  This contribution is often denoted by $\chi_{DOS}=\chi_{GG}$ . 
\cite{RandVarl,VarlReview,Carreta1,Eschrig} The second-order 
(Aslamazov-Larkin) diagram does not contribute to transverse susceptibility. 
 
We define  the static fluctuation propagator $\cal{D}$ in terms of  the particle-particle 
t-matrix as
\begin{equation}
{\cal D}({\bf k})  =   - k_{B} T \, t({\bf k}, 0)\, ,
\end{equation}
where
\begin{equation}
t^{-1}({\bf k},i\nu_{m}) = V^{-1} - k_{B} T \sum_{i\omega_{n}}\frac{d^{3}k'}
{(2\pi)^{3}} G({\bf k'},i\omega_n)G({\bf {q-k'}},i\nu_m-i\omega_n)\,\gamma \,.
\label{tmin11}
\end{equation}
Here $\omega_{n}=\omega_{n}(T)=(2n+1)\pi k_{B}T$, k is the total momentum 
of the Cooper pair, $\gamma$  is the impurity vertex,  and we have taken the 
pair interaction V to be constant.  ${\cal D}(k)$diverges when $T$ approaches 
$T_C$ from above for k equal to zero,  signaling the transition to the 
superconducting phase and defining $T_C$.  In order to describe static 
Gaussian fluctuations, the t-matrix is calculated for small total momentum 
with zero order Green's functions.  A standard weak coupling calculation 
\cite{Ambegaokar} yields for a clean system ($\gamma = 1$)		
\begin{equation}
{\cal D}^{-1}(k) = \left(\frac{N_{0}\pi}{k_{B}T}
\right) \left\{\sum_{\omega_{n}(T_C)}\frac{k_{B}T_C}{|\omega_{n}(T_C)|} - 
\sum_{\omega_{n}(T)}\frac{k_{B}T}{|\omega_{n}(T)|} + 
\frac{(\hbar v_{F} k)^2}{12}
\sum_{\omega_{n}(T)}\frac{k_{B}T}{|\omega_{n}(T)|^3}\right\}.
\label{Dmin11}
\end{equation}
where $N_{0}=ma^{2}/(2\pi\hbar^2)$   is the 2D density of states per spin and
$v_F$ is the Fermi velocity. All the sums are cut off at the constant BCS cutoff
$\omega_C$. The cutoff on $n$, $n_c=\frac{\omega_c}{2\pi T}-\frac{1}{2}$, 
depends on $T$ and is responsible for the important difference in the first two
sums in Eq. (\ref{Dmin11}). \cite{Allen} Performing the sums leads to
\begin{equation}
{\cal D}(k) = \frac{k_{B}T/N_0}{\ln(T/T_{C})+\xi^{2}k^2}\,,
\end{equation}
where $\xi$ is related to the BCS coherence length $\xi_{0} = \xi(T=0)$,
\begin{equation}
\xi = \sqrt{\frac{7\zeta(3)}{48}}\frac{\hbar v_F}{\pi k_{B}T} =
\xi_{0}T_{C}/T\, ,
\end{equation}
In the early conductivity calculations it was first observed that the 
Maki diagram has anomalous properties:  In the presence of non-magnetic 
impurity scattering, the impurity vertex corrections at each end of the 
fluctuation propagator $\cal{D}$ in the polarization bubble lead to a divergence 
at all temperatures in 2D.  This divergence could be removed by including a pair 
breaking parameter. \cite{Thompson,Keller} We emphasize that we work in the 
clean limit where there is no impurity vertex in the first place and thus this 
``Maki-divergence" does not occur.  As we will see however,  this contribution 
is also rather pathological in a clean system in 2D:  We find that it diverges at 
all temperatures when $\omega$  goes to zero.  This divergence is removed 
(except right at $T = T_C$) by inclusion of pair breaking.
  
The Maki-Thompson contribution to $T_1^{-1}$, $\chi_{FF}$,  was first evaluated 
in 2D by Kuboki and Fukuyama \cite{Kuboki} in the dirty limit for $\omega = 0$. 
They found a logarithmic divergence on approaching $T_C$ from above with 
the form $\ln(\Gamma/t)$    where $\Gamma$  is a pair breaking parameter.  
This presented a problem in the high-$T_C$ superconductors because no 
corresponding peak has been observed there.  An attempt to remedy this 
situation was made by Heym \cite{Heym} who generalized Kuboki and 
Fukuyama´s work by including the frequency dependence of  $\cal{D}$.  He 
found that dynamic fluctuations becomes important for  $T/T_{C} \ge$ 1.05  
and can lead to a significant correction to the $T$-dependence of $1/T_1$  
for these temperatures,  in particular for strong pair breaking.  The large peak 
for $T \rightarrow T_{C}$ remains however. We will see later that pair breaking 
can strongly reduce this peak.

We follow the notation of Maniv and Alexander \cite{ManivAlex} and write in 2D
\begin{equation}
\chi_{FF}(i\Omega_{\nu})=\sum_{\bf k}C({\bf k},i\Omega_{\nu}){\cal D}(k)\,,
\end{equation}
\begin{equation}
\chi_{GG}(i\Omega_{\nu})=\sum_{\bf k}A({\bf k},i\Omega_{\nu}){\cal D}(k)\,,
\end{equation}
where the contributions of the polarization diagrams are given by
\begin{eqnarray}
C({\bf k},i\Omega_{\nu}) 
& = &
 -k_BT\sum_{n}\sum_{{\bf p}_{1},{\bf p}_{2}}
G({\bf p}_{1},i\omega_{n})G({\bf k}-{\bf p}_{1},-i\omega_{n})
\nonumber\\
& \times &
G({\bf p}_{2},i\omega_{n}+i\Omega_{\nu}) 
G({\bf k}-{\bf p}_{2},-i\omega_{n}-i\Omega_{\nu})
\label{Cofk1}
\end{eqnarray}
and
\begin{equation}
A({\bf k},i\Omega_{\nu})=2k_{B}T\sum_{n}\sum_{{\bf p}_{1},{\bf p}_{2}}
G^{2}({\bf p}_{1},i\omega_{n})G({\bf k}-{\bf p}_{1},-i\omega_{n})
G({\bf p}_{2},i\omega_{n}+i\Omega_{\nu})\, .
\label{Aofk1}
\end{equation}
The G's are taken to be zero order propagators of the form (for the case 
of no pair breaking) $G({\bf p},i\omega_{n})=(i\omega_{n}-\xi_{\bf p})^{-1}$ 
where we assume free particles in 2D, $\xi_{\bf p}=p^2/2m - E_F$  . 
Since ${\cal D}(k)$ is peaked at small $k$, we approxoimate 
$\xi_{{\bf k}-{\bf p}}\approx\xi_{\bf p}-v_Fk\cos\phi$  where 
$\phi=\angle(\bf k,\bf p)$.  The factor 2 in  (\ref{Aofk1}) accounts for the 
two diagrams of the GG type.%
\vspace{0.5in}  

\subsection{Clean System without Pair Breaking}

We consider first the Maki-Thompson diagram. Performing the integrals 
over $\xi_{\bf p_1}$ and $\xi_{\bf p_2}$ in Eq. (\ref{Cofk1}) we have
\begin{eqnarray}
C({\bf k},i\Omega_{\nu})
& = &
N_{0}^2\pi^2T\sum_{n}sgn(\omega_n)
sgn(\omega_n+\Omega_{\nu})\nonumber\\
& \times &
\int_{0}^{2\pi}\frac{d\phi_1}{2\pi} 
\int_{0}^{2\pi}\frac{d\phi_2}{2\pi}
\frac{1}{(i\omega_{n}-v_{1}k/2)(i\omega_{n}+i\Omega_{\nu}-v_{2}k/2)}\, .
\end{eqnarray}
Due to the $sgn$ functions it is mathematically convenient to follow Maniv and
Alexander \cite{ManivAlex} and split the contribution into two parts:
\begin{equation}
\mbox{Im}\chi_{FF}(\omega)=\mbox{Im}\chi_{FF}^{(1)}(\omega) + 
\mbox{Im}\chi_{FF}^{(2)}(\omega)\, ,
\end{equation}
where
\begin{equation}
\mbox{Im}\chi_{FF}^{(j)}(i\Omega_{\nu}) =N_{0}^{2}\sum_{\bf k}{\cal D}(k)
\int_{0}^{2\pi}\frac{d\phi_1}{2\pi} \int_{0}^{2\pi}\frac{d\phi_2}{2\pi}
\Pi_{j}(\varepsilon_1,\varepsilon_2,i\Omega_{\nu})
\label{ImFFj}
\end{equation}
with  $\varepsilon_i = kv_F\cos\phi_i/2$, $j = 1,2$ and
\begin{equation}
\Pi_{1}(\varepsilon_1,\varepsilon_2,i\Omega_{\nu}) =
\pi^2k_BT\sum_{n=-\infty}^{+\infty}
\frac{1}{(i\omega_{n}-\varepsilon_{1})
(i\omega_{n} - \varepsilon_{2}+i\Omega_{\nu})}
\label{Pi1}
\end{equation}
\begin{equation}
\Pi_{2}(\varepsilon_1,\varepsilon_2,i\Omega_{\nu}) = -2\pi^2k_BT
\cases{ 
\sum_{n=-\nu}^{-1}\frac{1}{(i\omega_{n}-\varepsilon_{1})
(i\omega_{n} - \varepsilon_{2}+i\Omega_{\nu})} 
 &$\nu > 0$ \cr
\sum_{n=0}^{|\nu|-1}\frac{1}{(i\omega_{n}-\varepsilon_{1})
(i\omega_{n} - \varepsilon_{2}+i\Omega_{\nu})}
 &$\nu < 0$ \cr
}
\label{Pi2}
\end{equation}
For a clean system this decomposition is done merely for mathematical 
convenience. \cite{ManivAlex}  When impurity scattering is present however,  
the singular q dependence due to the impurity vertices at the ends of the 
$\cal{D}$ propagator (the ``Maki divergence",  again) makes it necessary to 
split $\chi_{FF}$ in a different manner into a "regular" and an "anomalous" 
part. \cite{RandVarl}  
Performing the sum over n in Eq. (\ref{Pi1})  and analytically continuing 
$i\Omega_{\nu}$  to the real external frequency $\omega$ ,  we find,  to 
leading order in $\omega$
\begin{equation}
\mbox{Im}\chi_{FF}^{(1)}(\omega) = \left[ \left(\frac{a}{\xi_0}\right)^2
\frac{\hbar\omega}{E_F}\frac{N_0}{2}\right]
\int_{\overline{\Omega}}^{\Lambda}{\cal D}(x)dx
\int_{-1}^{1-2\overline{\Omega}/x} \frac{dy}{\sqrt{1-y^2}}
\frac{(-1)f'(\varepsilon_c xy/\overline{T})}
{   \sqrt{1-\left( \frac{2\overline{\Omega}}{x}+y\right)^2}   }\enspace ,
\label{ImFF11}
\end{equation}
\begin{equation}
{\cal D}(x)=\frac{N_0\zeta^2}{k_B T}{\cal D}(k)=
\frac{\overline{T}^2}{x^2 + t(\overline{T}/\zeta)^2}\enspace ,
\label{Dofx}
\end{equation}
where we have defined  $x = k / k_F$,
$ \overline{\Omega}=\hbar\omega/2E_F$ , $\varepsilon_c=E_F/k_BT_C$ , 
$\overline{T}=T/T_C$, $\zeta=k_F\xi_0$ and 
$t=\ln\overline{T}\approx(T-T_C)/T_C$.  Here $f'$  is the derivative of the 
Fermi function and $\Lambda$  is a cutoff which we take as O(1).  A larger 
cutoff does not change the results significantly.  Note that, since $-f'$ is 
positive,  $\mbox{Im}\chi_{FF}^{(1)}$ is a positive contribution, proportional 
to $N_0^2$.

After analytic continuation and expansion to first order in $\omega$  we find 
from Eqs. (\ref{ImFFj}) and (\ref{Pi2})
\begin{eqnarray}
\mbox{Im}\chi_{FF}^{(2)}(\omega)
& = &
\left[ \left(\frac{a}{\xi_0}\right)^2
\frac{\hbar\omega}{E_F}\frac{N_0}{2}\right]
\left[\frac{\varepsilon_c}{\pi\overline{T}}\right]
\int_{0}^{\Lambda}x{\cal D}(x)dx\nonumber\\
& \times &
\int_{0}^{2\pi}d\phi_1 
\int_{0}^{2\pi}d\phi_2
\sum_{n=0}^{\infty}\left\{\frac{p_n\mu_1^2\mu_2^2 -p_n^5}
{(p_n^2 + \mu_1^2)^2(p_n^2 + \mu_2^2)^2}\right\}\, ,
\label{ImFF2}
\end{eqnarray}
where $\mu_i=\varepsilon_c\,x\cos\phi_i/\overline{T}$ , $p_n=(2n+1)\pi$, 
and several terms that vanish after the angular integrations have been 
omitted.  $\mbox{Im}\chi_{FF}^{(2)}$  is negative.

We turn now to the calculation of the DOS contribution $\chi_{GG}$ and first 
evaluate the sum over n in Eq. (\ref{Aofk1}) as a contour integration and then 
carry out the analytic continuation 
$i\Omega_{nu} \rightarrow \omega + i\delta$ .  Of the 
several resulting terms only the following yields, after the angular 
integrations, a non-vanishing contribution to first order in $\omega$ :
\begin{equation}
A({\bf k},\omega) = 4 \pi N_0\sum_{\bf p_1}\int_{-\infty}^{+\infty}
\frac{-f(\xi_2)d\xi_2}{(\xi_2-\xi_1-\omega-i\delta)^2
(\xi_2+\xi_1-{\bf v}\cdot{\bf k}-\omega-i\delta)}\,.
\end{equation}
The   $\xi_2$ integral can be expressed as a sum of the residues at the poles 
of the Fermi function:
\begin{equation}
A({\bf k},\omega) = 8\pi^2iN_0k_BT\sum_{\bf p_1}\sum_{n=-1}^{-\infty}
\frac{1}{(i\omega_n-\xi_1-\omega)^2
(\i\omega_n+\xi_1-{\bf v}\cdot{\bf k}-\omega)}\, .
\end{equation}
Expanding to first order in $\omega$  and performing the $\xi_1$  integration 
we obtain the negative contribution
\begin{equation}
\mbox{Im}\chi_{GG}(\omega) = -\left[ \left(\frac{a}{\xi_0}\right)^2
\frac{\hbar\omega}{E_F}\frac{N_0}{2}\right]
\left[\frac{8\varepsilon_c}{\pi\overline{T}}\right]
\int_{0}^{\Lambda}x{\cal D}(x)dx\int_{0}^{\pi/2}d\phi 
\sum_{n=0}^{\infty}
\left\{ \frac{p_n\left[p_n^2-3(x\varepsilon_c\cos\phi/\overline{T})^2\right]}
{\left[p_n^2+(x\varepsilon_c\cos\phi/\overline{T})^2\right]^3} \right\}
\quad .
\label{ImGG1}
\end{equation}
In Fig. 3 we plot the three contributions to $\mbox{Im}\chi$  as functions of 
the reduced Temperature $\overline{T}=T/T_C$ .  The positive contribution 
$\mbox{Im}\chi_{FF}^{(1)}$  is seen to dominate strongly.

It is interesting to consider these results in the limit of $T \rightarrow T_C$ 
from above. The experimental external frequency $\omega$  is very small 
but we leave it finite since $\mbox{Im}\chi_{FF}$  depends on the ratio 
$t/\omega$. In Eqs. (\ref{ImFF2}) and (\ref{ImGG1}) the function 
$x{\cal D}(x) \propto x/(x^2 + t\,\overline{T}^2/\zeta^2)$  is peaked at small $x$ 
for $t$ small. The sum over n is only weakly dependent on x for small $x$ 
and it is a good first approximation to set $x = 0$ within the sum.  The $t$ 
dependence then arises solely from the $x$-integration over the 
fluctuation propagator and one easily finds 
\begin{equation}
\frac{\mbox{Im}\chi_{FF}^{(2)}}{\omega}\propto\ln t\, ,\quad
\frac{\mbox{Im}\chi_{GG}}{\omega}\propto\ln t\, ,\quad
\mbox{for }t\rightarrow0\,.
\label{lim1}
\end{equation}
These limiting functions are independent of $\omega$ .  In 
Eq. (\ref{ImFF11}) for $\mbox{Im}\chi_{FF}^{(1)}$  the limiting 
behavior is not just due to the fluctuation propagator but is 
strongly affected by the integral over $y$.  We find the 
following result:
\begin{equation}
\frac{\mbox{Im}\chi_{FF}^{(1)}}{\omega}\propto
\cases{
\frac{1}{\sqrt{t}}\ln{\left(\frac{E_F\sqrt{t}}{\omega}\right)}
&\mbox{for }$\frac{\omega}{E_Ft}\rightarrow 0$ \cr
\frac{1}{\omega}
&\mbox{for }$\frac{E_Ft}{\omega}\rightarrow 0$ \cr
}
\label{lim2}
\end{equation}
For comparison we give the corresponding results for small $t$ and 
$\omega$ for the clean 3D case: \cite{ManivAlex}
\begin{equation}
\frac{\mbox{Im}\chi_{FF}^{(1)}}{\omega}\propto
\ln\left(\frac{1}{t+c_1\omega^2}\right)\,,\quad\
\frac{\mbox{Im}\chi_{FF}^{(2)}}{\omega}\propto c_2\,,\quad
\frac{\mbox{Im}\chi_{GG}}{\omega}\propto-c_3\,,
\label{lim3D}
\end{equation}
where $c_1$, $c_2$, and $c_3$ are positive constants. In comparison to 
3D, the results in 2D are rather pathological.  In 3D the only divergence 
occurs in  $\mbox{Im}\chi_{FF}^{(1)}/\omega$   and then only when 
{\it {both}} $t$ and $\omega$ go to zero.  In 2D, 
$\mbox{Im}\chi_{FF}^{(2)}/\omega$  and 
$\mbox{Im}\chi_{GG}/\omega$   diverge only for $t\rightarrow0$  ,  
while $\mbox{Im}\chi_{FF}^{(1)}/\omega$   diverges as 
$\omega\rightarrow0$  {\it {for all}} $t\,$!  This strange behavior is 
presumably unphysical and is removed by a small amount of pair breaking.  
As seen from Eq. (\ref{lim2}),  the limiting value of 
$\mbox{Im}\chi_{FF}^{(1)}/\omega$  depends on the ratio 
$\omega/t$ .  As  $t\rightarrow0$ there is a "crossover" near 
$t=\omega/E_F$  which is probably too close to $t = 0$ to be 
experimentally observable.  In any case,  for finite $\omega$  there is no 
divergence in $\mbox{Im}\chi_{FF}^{(1)}/\omega$   for 
$t\rightarrow0$ .

It is also interesting to compare our Eq. (\ref{lim2}) with the corresponding 
result of Randeria and Varlamov (RV) \cite{RandVarl} who give a result for 
the "ultra-clean" limit of a theory that includes elastic scattering from the 
beginning.  Their Eq.(16) reads
\begin{equation}
\mbox{Im}\chi_{FF}\propto\frac{1}{\sqrt{t}}\ln\left(T_C\,\tau\sqrt{t}\right)
\qquad\mbox{for   }T_C\,\tau\sqrt{t}\gg1\quad\mbox{and   }\omega=0\,,
\label{RV16}
\end{equation}
where $\tau$  is the lifetime for elastic scattering.  Since this equation does 
not contain their pair breaking parameter $\delta$,  it should be valid for 
the case of no pair breaking.  Equations (\ref{lim2}) and (\ref{RV16}) are 
similar in several respects.  In both cases there is a prefactor $1/\sqrt{t}$  
multiplied by the $ln$ of a large number proportional to $\sqrt{t}$  .  Since 
RV set $\omega=0$  before the calculation while we have $1/\tau=0$ ,  an 
exact comparison of the results is not possible.  In the exact clean limit 
Eq. (\ref{lim2}) seems preferable because  $\omega$ ,  although very 
small,  is a well defined experimental quantity while $\tau$  in 
Eq. (\ref{RV16}) is not well known and , more importantly, $\tau$  should 
not even appear in the exact result for a clean system.  Also,  the limit 
$t\rightarrow0$  is only possible if $\omega$ remains non-zero.

\subsection{Clean system with pair breaking}

	With "pair breaking" we mean essentially the effects of inelastic 
scattering which,  in the high $T_C$ superconductors, for example,  could be 
due to spin fluctuations and phonons.  In order to obtain a rough estimate of 
this effect we want to simulate it in a simple manner within a weak coupling 
theory.  Our procedure is equivalent to adding a constant inelastic scattering 
rate to the single particle propagator as done by other authors. \cite{Eschrig}  
We attempt to justify this physically by assuming that the effect on the 
fluctuation propagator $\cal{D}$ is similar to the well known pair breaking effect 
of scattering by magnetic impurities. \cite{Allen}  Impurity scattering affects 
$\cal{D}$ in two ways:  self energy corrections to the single particle 
propagators of the t-matrix ladder and vertex corrections to the pair interaction.  
For non-magnetic impurities these two contributions cancel for s-wave pairing 
(Anderson's theorem).  For magnetic impurities there is no cancellation and 
the transition temperature is strongly reduced.  For our purposes it is 
sufficient to retain only the self energy corrections and to assume that the 
vertex correction is included in the effective pairing interaction.  The 
Green's functions in Eq. (\ref{tmin11}) thus have the form
\begin{equation}
G({\bf k},i\omega_n)=\frac{1}{i\widetilde{\omega}_n-\xi_{\bf k}}
\end{equation}
with
\begin{equation}
\widetilde{\omega}_n=\omega_n+\Gamma {\text{sgn}}(\omega_n)\, ,
\label{Renorm}
\end{equation}
where $\Gamma \equiv \hbar/2\tau_{\phi}$  is the phenomenological pair 
breaking parameter destroying the phase coherence between the Cooper 
pairs.  Eq. (\ref{Dmin11}) for $\cal{D}$ is now replaced by
\begin{equation}
{\cal D}^{-1}(k) = \left(\frac{N_{0}\pi}{k_{B}T}
\right) \left\{\sum_{\omega_{n}(T_{C0})}
\frac{k_{B}T_{C0}}{|\omega_{n}(T_{C0})|} - 
\sum_{\omega_{n}(T)}\
\frac{k_{B}T}{|\widetilde{\omega}_{n}(T)|} + 
\frac{(\hbar v_{F} k)^2}{12}
\sum_{\omega_{n}(T)}
\frac{k_{B}T}{|\widetilde{\omega}_{n}(T)|^3}\right\},
\label{Dmin12}
\end{equation}
where $T_{C0}$ is the transition temperature in the absence of pair breaking.  
Although we report here only results for $\Gamma$ constant,  in general 
$\Gamma$ will be a function of $\omega_n$ .  For comparison, 
for $\Gamma$  constant the sums can be carried out analytically and 
Eq. (\ref{Dmin12}) can be expressed in terms of the digamma and 
tetragamma functions in the notation of Eq. (\ref{Dofx}),  as
\begin{equation}
{\cal D}^{-1}(x) =\frac{1}{\zeta^2}\left[ \ln\left(\frac{T}{T_{C0}}\right)
- \Psi\left(\frac{1}{2}\right) + \Psi\left(\frac{1}{2}+\alpha\right)\right]
-\left(\frac{x^2}{14\zeta(3)\overline{T}^2}\right)
\Psi^{(2)}\left(\frac{1}{2}+\alpha\right)\, ,
\label{Dmin13}
\end{equation}
where 
\begin{equation}
\alpha=\frac{\Gamma}{2\pi k_B T}\, .
\end{equation}
The transition temperature in the presence of pair breaking, $T_C$ ,  is 
obtained by setting ${\cal D}(0)^{-1}$ equal to zero yielding the 
Abrikosov-Gorkov equation. \cite{AGD}  Since we assume the pair breaking 
is due to inelastic scattering, $T_{C0}$ is not experimentally accessible.  
Thinking for the moment of the high $T_C$ superconductors, in our numerical 
calculations we 
take $T_C = 100\, \mbox{K}$ as experimentally given and $T_{C0}$ will be 
determined by the damping parameter $\Gamma$ .  Although $T_C < T_{C0}$ ,  
as expected,  $\cal{D}$ as a function of $k$ can be modified by the pair breaking in 
such a way that, neglecting the effect of pair breaking in the rest of the 
susceptibility diagrams,  the fluctuation contributions to $T_1^{-1}$ are increased.  
To elucidate this behavoir we show $\cal{D}$ as a function of $x = k / k_F$  in 
Fig. 4 for $T = 1.03\, T_C$ and several values of $\alpha_C = \alpha\,\overline{T}$.  
The figure shows that the maximum and width of ${\cal D}(k)$ increase with 
$\Gamma$.  The fact that the fluctuation propagator $\cal{D}$ increases with 
increasing pair breaking may seem counter-intuitive if one assumes, in analogy 
to the case of magnetic impurities where $T_{C0}$ is known, that:  pair breaking 
$\longrightarrow$  reduced tendency to superconductivity $\longrightarrow$ 
weaker fluctuations.  In that case,  for $T$ fixed relative to $T_{C0}$,  the 
fluctuations weaken with increasing $\Gamma$  because $T_C$,  and thus 
the divergence of $\cal{D}$,  are moving away from the reference point $T$.  
In our case, however,  $T$ is fixed relative to $T_C$ and $T_{C0}$ moves away 
with increasing $\Gamma$.  Thus the fluctuations at $T$ would be expected 
to be rather independent of $\Gamma$ .  That $\cal{D}$ actually increases arises 
mathematically from the dependence of the $ln$ term in Eq. (\ref{Dmin13}) on 
$\Gamma$  in this case.  The effect of pairbreaking in the single particle 
propagators however leads to a decrease of the susceptibility diagrams 
which usually dominates over the effect arising from $\cal{D}$.
			
We consider first $\chi_{FF}^{(1)}$  since it is the dominant contribution 
without pair breaking and its limiting $t$-dependence is 
qualitatively changed by the addition of pair breaking.  Equations 
(\ref{ImFFj}) and (\ref{Pi1}) are still valid if the frequencies are 
renormalized according to Eq. (\ref{Renorm}). The frequency sum and 
then the angular integrations can be carried out exactly with the result:
\begin{equation}
\mbox{Im}\chi_{FF\Gamma}^{(1)}(\omega) = \left[ \left(\frac{a}{\xi_0}\right)^2
\frac{\hbar\omega}{E_F}\frac{N_0}{2}\right]\, I_{\Gamma}
\label{FF1PB}
\end{equation}
\begin{equation}
I_{\Gamma}=\frac{\varepsilon_c}{16\pi^4\overline{T}\Omega}
\int_0^{\Lambda}dx\,x\,{\cal D}(x)
\int_{-\infty}^{+\infty}du\, f(2\pi\,u)\,Y(u,\eta,\alpha)
\left[Y(u+\Omega,\eta,\alpha)-Y(u-\Omega,\eta,\alpha)\right]\, ,
\label{IGam}
\end{equation}
where f is the Fermi function, $\eta = \varepsilon_c x/2\pi \overline{T}$,   
$\Omega = \hbar\omega / 2\pi k_B T$,  and 
$Y(a,b,c)=\pi \sqrt{2 \left[ \left( b^2-a^2+c^2+\sqrt{N}\right)/N\right]^{1/2}}$, 
with $N=\left(b^2-a^2+c^2\right)^2 + 4a^2c^2\,.$  We have computed $I_\Gamma$ 
numerically and the result as a function of the pair breaking parameter 
$\alpha_C$  is shown in Fig. 5 for several values of the external frequency, 
$\overline{\Omega}=\hbar\omega / 2E_F$ .  Note the very large and rapid 
change at small pair breaking for the experimentally relevant frequency\\ 
$\overline{\Omega}=5\times 10^{-8}$ .
						
Caculation of $\mbox{Im}\chi_{FF}^{(2)}(\omega)$  for finite $\Gamma$  yields
\begin{eqnarray}
\mbox{Im}\chi_{FF,\Gamma}^{(2)}(\omega)
& = &
\left[ \left(\frac{a}{\xi_0}\right)^2
\frac{\hbar\omega}{E_F}\frac{N_0}{2}\right]
\left[\frac{\varepsilon_c}{\pi\overline{T}}\right]
\int_{0}^{\Lambda}x{\cal D}(x)dx\int_{0}^{2\pi}d\phi_1 
\int_{0}^{2\pi}d\phi_2\nonumber\\
& \times &
\sum_{n=0}^{\infty}\frac{1}{(\mu_2-\mu_1)^2 +(4\pi\alpha_C)^2}
\left\{\frac{\left[ 2\pi \alpha_C\left(\mu_1^2+p_{n+}^2 \right) +
(\mu_2-\mu_1)\mu_1p_{n+}\right]}{\left( p_{n+}^2+\mu_1^2\right)^2 }
\right .
\nonumber\\
& - &
\left .
\frac{\left[ 2\pi \alpha_C\left(\mu_2^2+p_{n-}^2 \right) +
(\mu_2-\mu_1)\mu_2p_{n-}\right]}{\left( p_{n-}^2+\mu_2^2\right)^2}
\right\} \, , 
\label{FF2PB}
\end{eqnarray}
where $p_{n\pm}=p_n \pm 2\pi\alpha_C / \overline{T}\,$.  A similar 
calculation leads to the result that 
$\mbox{Im}\chi_{GG,\Gamma}(\omega)$  is simply given 
by Eq. (\ref{ImGG1}) with the replacement 
$p_{n}\rightarrow p_n + 2\pi\alpha_C / \overline{T}\,$ .

The limiting behavior of $\mbox{Im}\chi_{GG,\Gamma}(\omega)$  and  
$\mbox{Im}\chi_{FF,\Gamma}^{(2)}(\omega)$ for $t=(T-T_C)/T_C\rightarrow 0$  
is still given by Eq. (\ref{lim1}).  Similarly,  for finite $\Gamma$ ,  the 
T-dependence of $\mbox{Im}\chi_{FF,\Gamma}^{(1)}(\omega)$  now also 
arises primarily from the integral of $x{\cal D}(x)$ in Eq. (\ref{IGam}) and one 
finds $\mbox{Im}\chi_{FF,\Gamma}^{(1)}(\omega)\propto -\ln{}t$, for 
$t\rightarrow 0$. Thus,  in the presence of pair breaking,  the 
magnitudes of all three contributions have the same limiting 
Temperature dependence.

In Fig. 6 we plot the fluctuation contribution to
$\left[1/(T_1T)\right]_{\mbox{FL}}$, 
$\mbox{Im}\chi_\Gamma(\omega)$, given by equations   
 (\ref{T1Def}) and (\ref{Def2}), vs $\alpha_C$.
Note that $\mbox{Im}\chi_{FF,\Gamma}^{(2)}$ increases in magnitude with
increasing $\alpha_C$ until it dominates over 
$\mbox{Im}\chi_{FF,\Gamma}^{(1)}$ leading to a  
change of sign of the total contribution (dotted line) for the pair
breaking parameter $\alpha_C$ near  0.05.  This effect also occurs in the 
presence of weak elastic scattering. \cite{Eschrig}  In Fig. 7a we plot the 
total contribution vs $T/T_C$ for a range of $\alpha_C$ from 0 to 0.1.
The detailed plots in Figs. 7b and 7c, for small and large $\alpha_C$, 
show the dominance of  $\mbox{Im}\chi_{FF,\Gamma}^{(2)}$  for large pair 
breaking yielding a negative divergence.   We point out that,  in the 
presense of inelastic scattering (pair breaking),  our results here in the 
exact clean limit are qualitatively similar to those with weak elastic 
scattering. \cite{Eschrig}

To summarize briefly, we have shown that the Maki diagram in 2D is 
pathological in the exact clean limit but the divergences occur for a 
different reason than in the usual dirty limit. In the presence of pair
breaking due to inelastic scattering, which of course will always be 
present to some extent, reasonable results are obtained. A small
amount of pair breaking also strongly reduces the increase in 
$T_1^{-1}$ as $T_C$ is approached.  We have also seen that pair 
breaking in the fluctuation propagator ${\cal D}$ can affect $T_1^{-1}$
quite differently than pair breaking in the Green's functions in the
remainder of the diagram. 
In $\mbox{Im}\chi_{FF,\Gamma}^{(1)}(\omega)$ , for example, the 
increase in $T_1^{-1}$ due to pair breaking in ${\cal D}$ is more than
compensated by the decrease arising from the Green's functions.

The strong effect of even a very small amount of pair breaking 
within our simple weak coupling model underlines the need for a 
strong coupling Eliashberg calculation including inelastic 
scattering due to Boson exchange before quantitative comparison
with experiment can be attempted for specific systems.

\section{\quad Effect of BKT Vortex-Antivortex Fluctuations on $T_1^{-1}$
and $T_2^{-1}$}

Up to now we have discussed the effect of Gaussian fluctuations of
the order parameter on the NMR relaxation rate $T_1^{-1}$ near the
mean field transition temperature $T_{C0}$ of a 2D
superconductor. The weakly coupled 2D layers of high-$T_C$
superconductors promote the formation of
topological excitations in the form of pancake vortex-antivortex (VA)
pairs associated with the
singular part of the phase field of the complex
order parameter. Neglecting Josephson coupling, these VA excitations
reduce the transition temperature to a
Berezinskii-Kosterlitz-Thouless \cite{BKT} (BKT) transition
temperature $T_{\text{BKT}}$, above which the bound pairs start to
break up into free vortices and antivortices. We now proceed to study
the effect of magnetic-field fluctuations caused by the translational
motions of vortices and antivortices on $T_1^{-1}$ and $T_2^{-1}$ in
the vicinity of $T_{\text{BKT}}$.

\subsection{Vortex Fluctuations in BKT Theory}

In thin superconducting films, BKT theory \cite{BKT} predicts that
below a transition temperature $T_{\text{BKT}}$ spontaneously created
pancake vortices and antivortices (VA's) are bound in pairs with zero
total magnetic flux and do not destroy the off-diagonal
quasi-long-range order. Above $T_{\text{BKT}}$, the large pairs break
up into free vortices and antivortices, which are responsible for the
dissipation of electrical currents, and quasi-long-range order is
lost. Around $T_{\text{BKT}}$ there is a vortex-antivortex fluctuation
regime. The time and distance behavior of these fluctuations affect
the transport properties, e.g., BKT behavior is clearly seen
at microwave frequencies \cite{Corson} and in the
DC current-voltage characteristics. This picture is
essentially unchanged in layered superconductors if the Josephson 
coupling is ignored. \cite{layerEM,CTlay} (Under "Josephson 
coupling" one understands {\it all} interlayer pair transitions
contributing to the Josephson current.)

In weakly coupled high-$T_c$ superconductors such as
Bi$_2$Sr$_2$CaCu$_2$O$_{8+\delta}$ (Bi-2212) the results are 
changed, since the pancake vortices in the layers become 
connected by Josephson vortices between the layers. 
\cite{layerJ,Friesen,Blatter,CTlay} The Josephson vortices 
lead to a linear term in the interaction of pancakes connected 
in this way. The general picture is the following \cite{Blatter,Glaz}:
The VA pairs start to unbind at a BKT temperature $T_{\text{BKT}}$,
but the linear interaction leads to confinement of the pairs
at a length scale $\Lambda$, called the Josephson length. The BKT
renormalization is cut off at pair sizes of the order of $\Lambda$.
Very close to $T_{\text{BKT}}$, where the BKT correlation length
$\xi_{\text{BKT}}$ exceeds $\Lambda$, the interlayer coupling
becomes important and the system shows three-dimensional
critical behavior.\cite{footnote} The true transition takes
place at a temperature $T_C$.
At a higher temperature, where the BKT correlation length
$\xi_{\text{BKT}}(T)$ falls below $\Lambda$, 2D fluctuations
become important again and remain essential up to $T_{C0}$, where
the local condensation energy vanishes. Critical fluctuations
shift $T_{C0}$ slighty downwards from the mean field value. Around
$T_C$ there is a narrow 3D critical regime in which fluctuating
vortex loops through more than a single layer are crucial. We
are considering weakly coupled high-$T_C$ superconductors in the
sense that this 3D critical region is much narrower than the
2D fluctuation regime, which is well described by BKT theory.
We therefore neglect the Josephson coupling and restrict
ourselves to the 2D fluctuation region.

In the following we are concerned with the effect of BKT VA
fluctuations on the spin-lattice relaxation rate $T_1^{-1}$
and the spin-spin relaxation rate $T_2^{-1}$. Up to now, the
experiments on BKT fluctuations consist mainly of flux-noise
measurements \cite{Rogers,expnoise} with frequencies of the order of
$\omega\sim 10^4 \mbox{ s}^{-1}$, whereas NMR and NQR spectroscopy
are characterized by frequencies of order $10^7$ to
$10^8 \mbox{ s}^{-1}$.

\subsection{Vortex Relaxation Mechanisms}

We discussed above the microscopic contributions to the relaxation
rates arising from superconducting quasi-particles in the regime
of Gaussian fluctuations. Now we proceed to study the macroscopic
contributions of fluctuations due to VA excitations.
The vortices cause contributions from both the quasi-particle
excitations in their normal cores and from the fluctuating magnetic
fields carried by them. These fluctuating fields interact directly
with the nuclear magnetic moments and cause spin flip-transitions.
The nuclear spins in the vortex cores and in the superconducting
regions can be brought into thermal equilibrium by cross relaxation
through simultaneous spin flips of neighboring nuclei, i.e., by spin
diffusion. In type II superconductors, the cross relaxation of the 
nuclear spins in the normal core with the nuclei not in the core
tends to be suppressed by the mismatch of the Zeeman energies
in an inhomogeneous magnetic field. \cite{GR} However, in high-$T_c$
superconductors in the absence of an external field
cross relaxation is apparently not suppressed by this effect.
The experimentally observed line width, of the order of
$T_2^{-1}$, is about $10^5\mbox{ s}^{-1}$. On the other hand, the
magnetic field variations due to the vortices and the corresponding
variation of Zeeman energies of the $^{63}$Cu nuclei lead to an
inhomogeneous line width which is much smaller, of the order of
$10^3\mbox{ s}^{-1}$. \cite{CTthesis} Hence the system is
homogeneous; the $^{63}$Cu nuclei can be considered as a single
system subject to different relaxation mechanisms.
Furthermore, the motion of the vortex cores, which visit many
nuclei, also homogenizes the spin system.

We now focus on the relaxation of nuclear spins caused by the
fluctuating magnetic fields of the VA pairs. Since a static
magnetic field is not easily taken into account in BKT theory,
effects of BKT vortex fluctuations are best studied in the
relaxation rates measured in nuclear {\it quadrupole\/}
resonance (NQR) experiments on $^{63}$Cu nuclei which have
spin $I=3/2$ and nuclear quadrupole moment
$Q=-0.157\cdot 10^{-24}\mbox{ cm}^2$. The local electric field
gradient is oriented in the $z$ direction and gives a quadrupolar
splitting that is much larger than the Zeeman splitting due to
the magnetic field of the vortices. The relaxation rates are
governed by the time dependent correlation functions of the
magnetic field, which are determined by the VA fluctuations.
The evaluation of these correlation functions for the diffusing
and recombining vortices is the main task. We then proceed
to discuss the NQR relaxation rates in terms of the
magnetic-field fluctuations.

\subsection{Magnetic-Field Correlation Functions}

The Redfield theory gives the nuclear spin relaxation rates in
terms of correlation functions of the fluctuating magnetic
field. \cite{CTthesis,Slichter2}
We are interested in the correlation functions for the
time-dependent local magnetic field in a given layer,
\begin{equation}
k_{\alpha\beta}(t) \equiv
  \overline{h_{n,\alpha}({\bf r},t) h_{n,\beta}({\bf r},0)}
  = \overline{h_{0,\alpha}(0,t) h_{0,\beta}(0,0)} ,
\label{CTk1}
\end{equation}
where $h_{n,\alpha}({\bf r},t)$ is the $\alpha$ component of the
magnetic field at point ${\bf r}$ and time $t$ in layer $n$. This
field originates from the VA pairs in all the layers and is
given by
\begin{eqnarray}
{\bf h}_0(0,t) & = & \sum_n \sum_{\nu=1}^N \Big[
  {\bf H}_{-n}(-{\bf r}_{-n,\nu+}(t)) \nonumber \\
& & {}- {\bf H}_{-n}(-{\bf r}_{-n,\nu-}(t)) \Big] ,
\end{eqnarray}
where ${\bf r}_{n,\nu+}(t)$ and ${\bf r}_{n,\nu-}(t)$ are the
positions of the vortex and antivortex of the $\nu$-th pair in
layer $n$ at time $t$, and ${\bf H}_n({\bf r})$ is the magnetic
field in layer $n$ at ${\bf r}$ of a single vortex centered at the
origin. The components of this field parallel and perpendicular
to the layers are \cite{vxfield,AZdipl}
\begin{eqnarray}
{\bf H}_{n\parallel}({\bf r}) & = &
  \frac{\phi_0 s\,{\bf r}}{4\pi\lambda_{ab}^2r^2}\,\text{sign}(n)
  \Bigg[\exp\!\left(-\frac{|n|s}{\lambda_{ab}}\right) \nonumber \\
& & {}- \frac{|n|s}{\sqrt{r^2+n^2s^2}} \exp\!\left(
    -\frac{\sqrt{r^2+n^2s^2}}{\lambda_{ab}}\right)\Bigg] , \\
H_{n,z}({\bf r}) & = & \frac{\phi_0 s}{4\pi\lambda_{ab}^2
    \sqrt{r^2+n^2s^2}}\, \exp\!\left(
    -\frac{\sqrt{r^2+n^2s^2}}{\lambda_{ab}}\right) .
\end{eqnarray}
Here $\phi_0=hc/2e$ is the flux quantum, $\lambda_{ab}$ is the
penetration depth inside the layer, and $s$ is the inter-layer
separation. Whereas the $z$ component is of short range in both
the in-plane and the $z$ direction, the in-plane component is
of short range only in the $z$ direction and falls off with $1/r$
in the plane. For the convenient evaluation of the relaxation rates
we will later use the Fourier transforms
\begin{equation}
{\bf H}_{n\parallel}({\bf k}) = i\,
  \frac{\phi_0 s\,{\bf k}}{4\pi\lambda_{ab}^2k^2}\,
  \exp\!\left(-|n|s \sqrt{k^2+\lambda_{ab}^{-2}}\right)
\label{CTHFou1}
\end{equation}
for $n\neq0$, ${\bf H}_{0\parallel}({\bf k}) = 0$, and
\begin{equation}
H_{n,z}({\bf k}) = \frac{\phi_0 s}{4\pi\lambda_{ab}^2}\,
  \frac{1}{\sqrt{k^2+\lambda_{ab}^{-2}}}\,
  \exp\!\left(-|n|s \sqrt{k^2+\lambda_{ab}^{-2}}\right) .
\label{CTHFou3}
\end{equation}
We take the correlations between the fields of the vortex
and the antivortex of the same pair into account but neglect
inter-pair correlations. This is a good approximation,
since the typical pair size is small compared with
the average distance between pairs below the transition
and even in a significant temperature range above it. \cite{CThigh}
In the following we are only interested in the diagonal
components of $k_{\alpha\beta}$. They can be written as
\begin{eqnarray}
k_{\alpha\alpha}(t) & = & \frac{2N}{F} \sum_n \int d^2r_+'
  d^2r_-' d^2r_+ d^2r_-\; H_{-n,\alpha}(-{\bf r}_+') \nonumber \\
& & \times \Big[H_{-n,\alpha}(-{\bf r}_+)
  - H_{-n,\alpha}(-{\bf r}_-)\Big] \nonumber \\
& & \times P({\bf r}_+',{\bf r}_-';{\bf r}_+,{\bf r}_-;t)\,
  f({\bf r}_+-{\bf r}_-) .
\label{CTcorrel}
\end{eqnarray}
Here, we assume the presence of $N$ vortices and $N$ antivortices
in each layer so that $N/F\equiv n$ is the vortex density
calculated as a function of temperature in Refs.~\onlinecite{CThigh}
and \onlinecite{CTthesis}. The function $f({\bf r})$ gives the
normalized size distribution of the pairs. The {\it diffusion
function\/} $P$ describes the motion of the pairs within
a layer. We now discuss the functions $P$ and $f$.

We assume diffusive motion of vortices but take the
intra-pair interaction into account.
The diffusion function $P$ is defined as follows:
$P({\bf r}_+',{\bf r}_-';{\bf r}_+,{\bf r}_-;t) d^2r_+' d^2r_-'$
is the probability of finding the vortex of a given pair in the
area $d^2r_+'$ about ${\bf r}_+'$ and the antivortex of the same
pair in $d^2r_-'$ about ${\bf r}_-'$ at time $t$, provided the
vortex was at ${\bf r}_+$ and the antivortex at ${\bf r}_-$
at $t=0$. The diffusion function $P$ is discussed 
in a recent paper by Timm \cite{CTnoise} in the context of
flux noise. Here we summarize the results. Assuming for the
moment that vortices and antivortices are unbound, $P$ is
simply a product of free diffusion functions for ${\bf r}_+$
and ${\bf r}_-$ with a diffusion constant $D$. It
can be rewritten as
\begin{eqnarray}
\lefteqn{
P\left({\bf R}'+\frac{{\bf r}'}{2},{\bf R}'-\frac{{\bf r}'}{2};
  {\bf R}+\frac{{\bf r}}{2},{\bf R}-\frac{{\bf r}}{2};t\right)
  } \nonumber \\
& & \quad
  = \frac1{2\pi Dt}\,\exp\!\left(-\frac{|{\bf R}'-{\bf R}|^2}
  {2Dt}\right)\, \nonumber \\
& & \qquad \times 
  \frac1{8\pi Dt}\,\exp\!\left(
  -\frac{|{\bf r}'-{\bf r}|^2}{8Dt}\right) ,
\label{CTPfree}
\end{eqnarray}
which is the product of
free diffusion functions for the center of mass ${\bf R}$ and the
separation vector ${\bf r}$ of a pair, showing that the center of
mass and the separation vector diffuse with $D_{\text{cm}}=D/2$ 
and $D_{\text{rel}}=2D$, respectively. The assumption of unbound 
pairs would be justified for a small density of essentially free 
vortices, a situation that does not arise in practice. We take into 
account the interaction between vortex and antivortex, which is 
logarithmic in distance $r$, $V(r)\approx q^2\ln (r/r_0)$, where $q$ 
is the charge of the vortex in the Coulomb gas model and $r_0$ 
can be chosen as the size of the vortex core, and solve the
Fokker-Planck diffusion equation. The result is
\begin{eqnarray}
\lefteqn{
P\left({\bf R}'+\frac{{\bf r}'}{2},{\bf R}'-\frac{{\bf r}'}{2};
  {\bf R}+\frac{{\bf r}}{2},{\bf R}-\frac{{\bf r}}{2};t\right)
  } \nonumber \\
& & \quad = \frac1{2\pi Dt}\,\exp\!\left(-\frac{|{\bf R}'-{\bf R}|^2}
    {2Dt}\right)\,P_{\text{rel}}({\bf r}',{\bf r};t) ,
\end{eqnarray}
where the first term accounts for the motion of the center of mass
and $P_{\text{rel}}$ gives the relative motion in polar
coordinates $r$, $\phi$,
\begin{eqnarray}
P_{\text{rel}} & = & \frac1{4\pi D_{\text{rel}}t}\,
  \left(\frac{r'}{r}\right)^{\!\gamma}
  \exp\!\left(-\frac{r^{\prime 2}+r^2}{4D_{\text{rel}}t}\right)
  \nonumber \\
& & \times \sum_{n=-\infty}^\infty e^{in(\phi'-\phi)}\,
  I_{\sqrt{\gamma^2+n^2}}\!\left(\frac{rr'}{2D_{\text{rel}}t}\right) .
\end{eqnarray}
Here, $\gamma=(1-q^2/k_BT)/2$ and $I_\mu(x)$ is a modified Bessel
function.  Note that $P$ incorporates the effect of pair
recombination: Pairs with zero separation are taken out of the
process. Thus $P$ starts out normalized to unity at $t=0$ but then
drops to zero on the time scale of the life time $\tau_r$ of a VA
pair, determined by $D_{\text{rel}}=2D$.  Newly created VA pairs 
do not enter in $k_{\alpha\beta}$, since
there positions are assumed to be uncorrelated to those of
existing pairs. If the diffusion of a vortex were limited
only by the Bardeen-Stephen friction mechanism \cite{BS} (no pinning)
with a large diffusion constant, $\tau_r$ would be much smaller
than the time scale of NQR spectroscopy, of order $10^{-7}$ to
$10^{-8}\mbox{ s}$. We will come back to this problem below.

Besides the diffusion function, the correlation function in
Eq.~(\ref{CTcorrel}) contains the distribution function of pair
sizes, $f(r)$. Taking into account that the magnetic field
of a vortex changes on the scale of the penetration depth
$\lambda_{ab}$, and that for this reason the fields of a vortex
and an antivortex almost cancel for separations $r\ll\lambda_{ab}$,
we approximate the pair distribution function with an analytical
expression that becomes exact for large pairs and does not introduce
irrelevant complications for small $r$,
\begin{equation}
f(r) \propto \frac{1-(r/r_0)^2}{1-(r/r_0)^{2\zeta+6}} ,
\end{equation}
which should be normalized to unity. The form of the
exponent $\zeta$ is given in Ref.~\onlinecite{CTnoise}. We 
only note that $\zeta$ vanishes for $T\ge T_c$ and is positive 
and, to leading order, proportional to $(T_c-T)^{1/2}$ below 
$T_c$, where we now denote the BKT transition
temperature by $T_c$.

\subsection{NQR Relaxation Rates}

We can now evaluate the NQR relaxation rates in terms of the
correlation functions, Eq.~(\ref{CTcorrel}). The NQR
rates are of interest here because the BKT vortex fluctuations
show up more clearly in the absence of Abrikosov vortices due to
an external magnetic field. In a vanishing field, however, the
occupation of the quadrupolar energy levels cannot be described
in terms of a Boltzmann distribution with a spin temperature and,
therefore, we proceed by calculating the NQR relaxation rates
using the Bloch-Wangsness-Redfield theory. \cite{BWR} We can apply
this theory, since the interaction between the nuclear spins
and the magnetic field of the vortices is a small perturbation
compared with the quadrupolar splitting. The energy levels of
in-plane $^{63}$Cu nuclei due to quadrupole splitting are
shown in Fig. 8. Since the field gradient is oriented
along the $z$ direction and $I=3/2$, the relaxation rates are
given by
\begin{eqnarray}
T_1^{-1} & = & \sqrt{\pi/2}\,\gamma_n^2\,\Big[
  9k_{xx}(0)-7k_{xx}(\omega)\Big] ,
\label{CTrate1} \\
T_2^{-1} & = & \sqrt{\pi/2}\,\gamma_n^2\,\left[
  \frac{1}{4} k_{xx}(\omega)+\frac{3}{4} k_{xx}(0)+k_{zz}(0)\right] ,
\label{CTrate2}
\end{eqnarray}
where $\gamma_n=7.1\mbox{ G$^{-1}$s$^{-1}$}$ is the gyromagnetic
ratio of the $^{63}$Cu nuclei and 
$\omega = |\omega_{+3/2}-\omega_{+1/2}| =
|\omega_{-3/2}-\omega_{-1/2}|$. The correlation function
given by the temporal Fourier transform of Eq.~(\ref{CTcorrel})
has the form
\widetext
\begin{eqnarray}
k_{\alpha\alpha}(\omega) & = & \frac{2N}{F}\,\frac{8\sqrt{2\pi}}
  {D_{\text{rel}}} \int d^2k \sum_n |H_{-n,\alpha}({\bf k})|^2
  \int_0^\infty \! dr\,r^{1-\gamma}\,f(r) \nonumber \\
& & \times \sum_{m=1,\text{ odd}}^\infty J_m(kr/2) \int_0^\infty
  \! dr'\,r^{\prime 1+\gamma} J_m(kr'/2) \nonumber \\
& & \times \text{Re } I_{\sqrt{\gamma^2+m^2}}\left(\sqrt{
  k^2/4+i\omega/D_{\text{rel}}}\,r_<\right)
  K_{\sqrt{\gamma^2+m^2}}\left(\sqrt{k^2/4+i\omega/D_{\text{rel}}}
  \,r_>\right) ,
\label{CTkaa}
\end{eqnarray}
where $J$, $K$, and $I$ are Bessel functions, $\alpha=x,y,z$,
$r_<=\min(r,r')$, and $r_>=\max(r,r')$. This equation shows how
the spectral density of the magnetic field fluctuations determines
the relaxation rates, Eqs.~(\ref{CTrate1}) and (\ref{CTrate2}).
With the Fourier transforms of the vortex magnetic field
substituted from Eqs.~(\ref{CTHFou1}) and (\ref{CTHFou3}), the final
form of the correlation functions is
\begin{eqnarray}
k_{xx}(\omega) & = & \frac{2N}{F}\,\frac{4\sqrt{2\pi}}{D_{\text{rel}}}
  \,\frac{\phi_0^2 s^2}{4\pi\lambda_{ab}^4} \int_0^\infty \frac{dk}{k}
  \,\frac1{\displaystyle\exp\left(2s\sqrt{k^2+\lambda_{ab}^{-2}}
    \right)-1} \int_0^\infty \!dr\,r^{1-\gamma}\,f(r) \nonumber \\
& & \times \sum_{m=1,\text{ odd}}^\infty J_m(kr/2) \int_0^\infty
  \! dr'\,r^{\prime 1+\gamma} J_m(kr'/2) \nonumber \\
& & \times \text{Re } I_{\sqrt{\gamma^2+m^2}}\left(\sqrt{
  k^2/4+i\omega/D_{\text{rel}}}\,r_<\right)
  K_{\sqrt{\gamma^2+m^2}}\left(\sqrt{k^2/4+i\omega/D_{\text{rel}}}
  \,r_>\right)
\label{CTkxx}
\end{eqnarray}
and $k_{zz}(\omega)$ follows with a similar form.
\narrowtext

The effect of motional narrowing is taken into account: The nuclei
are fixed and experience the field fluctuations of the vortices as they 
pass by. An important parameter for diffusive relaxation phenomena 
is the characteristic frequency $\omega_c$ corresponding
to the inverse jump time for the diffusion process. \cite{Torrey}
Here, $\omega_c\sim D_{\text{rel}}/4\lambda_{ab}^2$, the reason
being that the dependence on $\omega$ is determined by
$k^2/4+i\omega/D_{\text{rel}}$ and the characteristic value of
$k$ is $1/\lambda_{ab}$. By inspection of $k_{\alpha\alpha}(\omega)$
it is seen that the correlation function becomes really frequency 
dependent only for frequencies $\omega> \omega_c$. The
Fourier transform $k_{\alpha\alpha}(t)$, Eq.~(\ref{CTk1}),
begins to decrease for $t< 1/\omega_c$. However,
$k_{\alpha\alpha}(t)$ does not have the simple exponential decay
form characteristic for the case of nuclei diffusing in a static,
random magnetic field.

For the special case of unbound pairs, see Eq.~(\ref{CTPfree}), the
result for $T_1^{-1}$ takes the simple form \cite{AZdipl2}
\begin{eqnarray}
T_1^{-1} & = & \frac{3\gamma_n^2n_P\phi_0^2}{4\pi}\,
  \frac{D}{\omega^2\lambda_{\text{eff}}^2} \int_0^{1/\xi_{ab}} \!\!\!
  dk\: \frac{k}{1+D^2k^4/\omega^2} \nonumber \\
& & \times \left[\exp\left(2s\sqrt{k^2+\lambda_{ab}^{-2}}
  \right)-1\right]^{-1} ,
\label{CTrate0}
\end{eqnarray}
where $n_P$ is the temperature-dependent pair density,
$\lambda_{\text{eff}} = 2\lambda_{ab}^2/s$ is the effective
penetration depth, \cite{Blatter,Glaz} and $\xi_{ab}$ is the in-plane
coherence length. In the general case there is no such simple form.
We now proceed to discuss the results in terms of the diffusion
constant $D$.

\subsection{BKT Vortex Fluctuations, $T_1^{-1}$, and $T_2^{-1}$}

The most important parameter in Eq.~(\ref{CTkxx}) is the diffusion
constant $D_{\text{rel}}=2D$ that governs both the free motion
of independent vortices and the motions of a vortex and an
antivortex towards each other, leading eventually to their
recombination. In the absence of pinning, $D$ is given by the
Bardeen-Stephen formula \cite{BS}
\begin{equation}
D = D_0 \equiv \frac{2\pi c^2\xi_{ab}^2\rho_n k_BT}{\phi_0^2 d} ,
\label{CT.BS}
\end{equation}
where $\rho_n$ is the normal-state resistivity and $d$ is the
layer thickness. For Bi-2212, $D_0$ is of the order of 
$1\mbox{ cm}^2/\mbox{s}$ at low
temperatures. \cite{March} This value is so large that 
vortices and antivortices would
recombine so fast that there is no time left for a large
number of nuclear relaxation processes to occur.
Futhermore, also when ignoring recombination processes,
the free motion of vortices and antivortices is so fast that
the rapidly changing magnetic fields at the nuclei lead to
slow relaxation, similar to the case of motional narrowing.

Slow diffusion rates and correspondingly long VA recombination
times are crucial for obtaining measurable contributions to
$T_1^{-1}$ and $T_2^{-1}$. In real high-$T_c$ superconductors
there are always inhomogeneities, i.e., doping
defects or intrinsic crystalline defects (twin boundaries,
grain boundaries), which can pin vortices by their interaction
with the normal vortex cores. A measure of the strength of
the pinning potential is the energy $E_p$ for the thermally
activated motion of a vortex. Pinning leads to an
Arrhenius-type temperature dependence,
\begin{equation}
D = D_0 \exp\left(-\frac{E_p}{k_BT}\right) .
\label{CTDact}
\end{equation}
In terms of $D$ the mean time between flights, or the time
of stay, is given by
$\tau_p = \ell_{\text{eff}}^2/4D$ ,
where $\ell_{\text{eff}}$ is a measure of the flight
distance. \cite{Aprem0} Experimentally, $D$ has been determined for
different thicknesses of layered CuO$_2$ systems. For a one-unit-cell
thick film of Bi-2212 Rogers
{\it et al}. \cite{Rogers} measure the low-frequency flux
noise $S_\phi$ near $T_c$. By analyzing the frequency
dependence of $S_\phi$ in terms of diffusion noise, the
authors determine a characteristic frequency
$\omega_c = D(T)/2\langle r\rangle^2$, above which
$S_\phi\propto\omega^{-3/2}$; $\langle r\rangle$ is the average pair
size. From these experiments the authors determine $D(T)$,
Eq.~(\ref{CTDact}), and find a temperature-dependent activation
energy for a single CuO$_2$ layer,
$E_p(T) \approx E_0 (1-T/T_{c0})$,
with $E_0/k_B \approx 800\mbox{ K}$. For more than one layer thick
epitaxial blocks of DyBa$_2$Cu$_3$O$_{7-x}$, the activation energy
$E_p$ is proportional to the number of layers, $N$, and begins
to saturate at $N=3$ to $N=4$. \cite{Fabrega} This is taken as
direct evidence that the pancakes in two adjacent layers are coupled
due to the Josephson effect and move as entities. With $N$ up to
$10^5$, large activation energies $E_p\approx 8\mbox{ eV}$ 
(consisting of the nucleation energy plus the pinning energy) are
observed in epitaxial films of Bi-2212,
where a BKT transition is experimentally found in the temperature
dependence of the penetration depth. \cite{KK}
In such crystals, the BKT transition may be driven by thermally
created pairs of VA {\it lines\/} through several layers.

To sum up, the diffusion constant $D$ for a pancake vortex in a
single clean CuO$_2$ layer is large,
$D\sim 1\mbox{ cm$^2$/s}$ so that the VA life time is short.
However, in the real multi-layer systems where the BKT transition is
observed, \cite{KK,Aprem1} $D$ can be small and thereby enhance
the life time of a VA pair so that the relaxation of nuclear spins
can accompany the translational diffusion of these pairs. The
actual value of $D$ can vary over many orders of magnitude; for
this reason we treat $D$ as an open parameter in the following
discussion.

Let us first comment on $T_1^{-1}$ ignoring the interaction and
recombination of VA pairs, cf.~Eq.~(\ref{CTrate0}). We assume
a multilayer structure of alternating single layers of
YBa$_2$Cu$_3$O$_{7-x}$ and PrBa$_2$Cu$_3$O$_{7-x'}$; the distance
between the superconducting CuO$_2$ layers is $s=24\mbox{ \AA}$.
Using $\lambda_{ab}\approx 1400\mbox{ \AA}$,
$\xi_{ab}\approx 12\mbox{ \AA}$,
and taking the experimental value $D=2\cdot 10^{-4}\mbox{ cm$^2$/s}$,
measured by Fiory {\it et al} \cite{Fiory}, for a thin film of
YBa$_2$Cu$_3$O$_{7-x}$ near $T_c$, we get from Eq.~(\ref{CTrate0})
a rate of $T_1^{-1}\approx 572\mbox{ s}^{-1}$, which is comparable
with the experimental values.

Next, let us take into account the interaction and recombination
of vortices and antivortices. The time it takes for a vortex to
move to its nearest antivortex depends on the VA separation and on $D$
for the single-vortex motion. Since $D$ is not known, we evaluate
the relaxation rates $T_1^{-1}$ and $T_2^{-1}$ as functions of the
diffusion constant of the separation vector, $D_{\text{rel}}=2D$.
The curves shown in Fig. 9 are obtained by numerical
integrations from Eqs.~(\ref{CTrate1})--(\ref{CTkaa}) using
parameter values that apply to Bi-2212.
The magnetic penetration depth and the Ginzburg-Landau coherence
length are given by
$\lambda_{ab}(T)/\lambda_{ab}(0)
  = \xi_{ab}(T)/\xi_{ab}(0) = \sqrt{T_{c0}/(T_{c0}-T)}$,
where $\lambda_{ab}(0)\approx 2000\mbox{ \AA}$,
$\xi_{ab}(0)\approx 21.5\mbox{ \AA}$, and $T_{c0}\approx 86.8\mbox{ K}$
is the mean-field transition temperature. The distance between
layers is $s\approx 15.5\mbox{ \AA}$. The parameter
$\gamma=(1-q^2/k_BT)/2$ depends on the VA interaction,
$q^2=q_0^2(T_c-T)$, where $k_BT_c/q_0^2\approx 0.2053$
and for $T_c$ we take the value $84.7\mbox{ K}$. \cite{CTthesis}
Figure 9 shows that the prefactor $1/D_{\text{rel}}$ in
Eq.~(\ref{CTkaa}) dominates the dependence of both relaxation rates
on $D_{\text{rel}}$; both rates fall off approximately as
$1/D_{\text{rel}}$. The reason is that a large value of
$D_{\text{rel}}$ leads to fast recombination. \cite{Aprem2}
Then, the time-dependent correlation functions $k_{\alpha\alpha}(t)$
are narrow and their Fourier transforms $k_{\alpha\alpha}(\omega)$,
Eq.~(\ref{CTkaa}), are broad and quite small even at $\omega\approx0$.
Hence, small diffusion constants are necessary to observe contributions
from VA fluctuations to the relaxation rates.

The diffusion constant has an exponential temperature dependence if
there are pinning effects. The corresponding temperature dependences
of $T_1^{-1}$ and $T_2^{-1}$ are evaluated with $D$ from
Eq.~(\ref{CTDact}), using for $E_p(T)$ the experimental results of
Rogers {\it et al}. \cite{Rogers} The bare diffusion constant $D_0$
is given by Eq.~(\ref{CT.BS}), it depends on $T$ through $\xi_{ab}$
and $\rho_n$, the normal resistance of a single layer. The
experimental value of $D(T_c)$ is large, not much smaller than
$1\mbox{ cm$^2$/s}$. Assuming the experimental values of $D(T)$
obtained by Rogers {\it et al}., the calculated relaxation rates
in the vicinity of $T_c=84.7\mbox{ K}$ are shown in Figs. 10
and 11. It is found that the temperature dependences of both
rates are qualitatively similar to the quasi-particle contributions
above and below $T_c$. There is a sharp drop of the rates below $T_c$.
The vortex contribution to $T_2^{-1}$ is smaller than the contribution
to $T_1^{-1}$. Since the observed spin-spin relaxation rates are larger
than the spin-lattice relaxation rates, the latter are more suitable
for an experimental test.

The crucial point in our example is the small absolute values of the
calculated relaxation rates. The numbers obtained by assuming a large
value of $D$ are several orders of magnitude smaller than the
experimental values. These are of the order of $10^2$ to
$10^3\mbox{ s}^{-1}$ above $T_c$, where the relaxation is
predominantly caused by quasi-particle excitations. Furthermore,
we have mentioned above that quasi-particle excitations in the
cores will also contribute to the relaxation effects from vortices.
In a simple picture the vortex cores can be considered as
normal-conducting regions, where normal-state-like excitations
contribute to $T_1^{-1}$ according to the Korringa law
$(T_1T)^{-1} = K_0$; here $K_0$ is the Korringa constant in the
normal state. In the superconducting state the vortex-core
contribution to $T_1^{-1}$ is approximately given by
\begin{equation}
\frac1{T_{1,\text{core}}T} = 2n_P(T)\:\pi\xi_{ab}^2(T)\:K_0 ,
\end{equation}
where $2n_P$ is the areal density of vortices and $\pi\xi_{ab}^2$
is the area of one core. The calculated temperature dependence
of $T_{1,\text{core}}^{-1}$ for Bi-2212
is shown in Fig. 12. A very similar curve would result
for $T_{2,\text{core}}^{-1}$. The core contribution is much larger
than the contribution from the VA magnetic-field fluctuations,
Fig. 10.

There are, however, several effects that can make an experimental
observation of the effect of the magnetic-field fluctuations feasible.
We have already discussed possible mechanisms by which the diffusion
constant $D$ can become smaller, namely pinning and coherent motion
of stacks of vortices. Even if $D$ is sufficiently small,
however, the VA contribution to $T_1^{-1}$
and $T_2^{-1}$ is at best comparable with the quasi-particle
contributions. In order to evade the in-plane quasi-particle
effects, inter-plane ions can be used for NQR spectroscopy. The
relaxation of inter-plane nuclear moments can be affected by the
magnetic-field fluctuations originating from vortices and
antivortices in the CuO$_2$ planes. \cite{Pieper} A possible
candidate is $^{201}$Hg in HgBa$_2$CaCu$_2$O$_{6+\delta}$ with
the $^{201}$Hg ion with spin $I=3/2$ and abundance $13.2\%$
residing between two CuO$_2$ planes.
So far NMR spectroscopy has been carried out only on the $^{199}$Hg
nucleus with $I=1/2$; an NMR spin-lattice relaxation time of
$T_1=32\mbox{ ms}$ is found at room temperature. \cite{Hoffmann}
Although this value will increase with decreasing temperature
if Korringa's relation applies, the relaxation rate $T_1^{-1}$ at
$T_c$ could still be an order of magnitude larger than the contribution
from magnetic-field fluctuations. Other inter-planar candidates may
be $^{87}$Sr with $I=9/2$ and abundance $7\%$ and $^{209}$Bi
with $I=9/2$ and abundance $100\%$, both in Bi-2212. The isotope
$^{43}$Ca ($I=7/2$) is probably not a good candidate, since its
abundance is rather small. To extend our approach to inter-plane
ions, the sums over squared Fourier transforms of the vortex
magnetic field in Eq.~(\ref{CTkaa}) have to be taken over the
appropriate fractional values for $n$. There
are other inter-plane ions with $I>1/2$ that are candidates for
NQR relaxation measurements in order to observe the magnetic fields
caused by the diffusional motion of BKT vortices. A pertinent
question concerns the relative magnitudes of the magnetic fields
at the inter-layer sites that originate from VA fluctuations
and from the fluctuating Cu moments in the CuO$_2$ planes,
respectively.

\section{\quad Summary, Conclusions, and Comments on Experiments}

In this paper we studied the effect of 2D superconducting fluctuations
on the nuclear spin relaxation of layered superconductors above the
superconducting transition temperature $T_C$ and outside the
region where Josephson coupling between the layers leads to a
narrow 3D critical region surrounding $T_C$. Hence we model the
superconductor as a set of 2D layers without interlayer coupling
mechanisms that transfer Cooper pairs between adjacent layers. 
The fluctuations in each layer consist of the usual Gaussian 
fluctuations and the topological Berezinskii-Kosterlitz-Thouless 
\cite{BKT} (BKT) vortex-antivortex fluctuations. The Gaussian 
fluctuations are considered as the long wavelength amplitude 
fluctuations of the superconducting order parameter and the
BKT fluctuations are taken into account as the shorter 
wavelength phase fluctuations of each layer. These two
fluctuation effects dominate the superconducting fluctuation 
behavior in the 2D fluctuation regime (cf. Fig. 1) of layered
systems.

We first evaluated in Sec. II, for s-wave superconductors, the
effect of amplitude fluctuations on the hyperfine relaxation rate
$T_1^{-1}$ for pure systems, with and without pair breaking effects
due to inelastic scattering. The amplitude fluctuations affect the 
Pauli spin susceptibility $\chi$ of the itinerant charge carriers
and thereby, via the hyperfine coupling, the decay of the nuclear
polarization due to the spin-lattice relaxation. Our results for the
relative effect of fluctuations on $T_1^{-1}$ in zero magnetic field
depends strongly on the pair breaking parameter 
$\alpha_C = \hbar / 4\pi \tau_{\phi} k_B T_C$ as is seen from
Figs. 6 and 7.  The results for clean systems without pair breaking 
are given by Eqs. (\ref{lim1}) - (\ref{lim3D}) and with pair breaking 
by Eqs. (\ref{FF1PB}) - (\ref{FF2PB}). Assuming small pair breaking
effects, $\alpha_C\approx 0.01$, and the temperature just outside
the critical region above $T_C$, say $T > 1.01 T_C$, a fluctuation
enhancement of $T_1^{-1}$ occurs that is of order $10$ with a slow
decrease as the temperature moves away from $T_C$, see Fig. 7.

Let us comment on the feasibility of experimental verification of
this enhancement effect and its temperature dependence.  A difficulty 
can arise in observing this effect: The high frequency fields used
in NMR experiments on metals, and especially superconductors, 
is shielded within the penetration depth of the radio-frequency 
field. \cite{Pieper2} For this reason most of the experiments are
performed on powders with grain sizes smaller than the skin depth
(in metals) or the penetration depth (in superconductors), or on thin 
films. In powders or thin films, however, the NMR lines are 
broadened by charge  density fluctuations that emanate from crystal
surfaces, an effect similar to the Friedel oscillations of the charge
density surrounding an impurity in a simple metal. \cite{Char} This
geometrical broadening effect will primarily affect the relaxation
rate $T_2^{-1}$ for nuclear spin-spin coupling, i.e., the natural line 
width. The charge fluctuations can also affect $T_1^{-1}$ because
of the change of the electron density at E$_{\text{F}}$. Hence, the
singular behavior of $T_1^{-1}$ near the transition, seen in Fig. 7, 
may be smeared out by surface effects. \cite{Walstedt} Furthermore, 
the spread of c-axis orientations in powder samples can also smear 
the effect of fluctuations on $T_1^{-1}$. \cite{Carreta2}

At this point one must ask what the chances are of finding systems
in which these properties can be observed. Aside from possibily the 
electron-doped high-$T_C$ superconductors mentioned in the 
introduction, there are other quasi-2D  superconductors with s-wave
pairing and BCS behavior.  The relaxation time $T_1$ of the $^{93}$Nb
nuclear spins in a single crystal of 2H-NbSe$_2$ (with trigonal prismatic
coordination of the Nb atoms) has been measured by Wada. \cite{Wada}
The rate follows the Korringa relation in the normal state and increases
exponentially in the superconducting state. Other dichalcogenides include
the TaS$_{2-x}$Se$_x$ layer cpmpounds which are BCS superconductors
and can be intercalated with organic molecules to increase the interlayer 
separation to as much as 50 $\AA$. \cite{Smith} Finally there are the
graphite intercalation compounds such as C$_8$K which are BCS
superconductors with rather low $T_C$'s. \cite{Dress} One or the other
of such quasi-2D systems may be a suitable candidate for observing 
the $T_1$ fluctuation effects discussed in this paper. The experimentalist 
may prefer to measure $T_1$ on a nucleus without a quadrupolar moment 
in order to avoid the line-broadening effect caused by electric field
gradients near the metal surfaces of thin films or small particles.

In Sec. III we studied the effect of Berezinski-Kosterlitz-Thoules
vortex-antivortex fluctuations on the NQR spin-lattice and spin-spin 
relaxation rates, $T_1^{-1}$ and $T_2^{-1}$, respectively. Here the 
symmetry of the superconducting order parameter will affect the field
and current distributions of an individual vortex. However, for both 
s- and d-wave superconductors, the flux contained in a vortex is
$\Phi_0$ , the magnetic field not too close to the vortex core does
not depend on the symmetry of the order parameter, 
and, therefore, this symmetry will not
affect the magnetic field fluctuations caused by the translational 
motions of an ensemble of VA pairs. Hence, the results for the 
relaxation rates $T_1^{-1}$ and $T_2^{-1}$ calculated in Sec. III can be 
applied to both s- and d-wave superconductors. One must keep in
mind, however, that the relaxation processes of the VA pairs are not 
entirely the result of the translational diffusion processes. The relaxation 
processes due to quasi-particle excitations must also be taken into 
account and can be different for the two symmetries. We assume that 
clear experimental evidence exists for the BKT transition in 
quasi-two-dimensional systems, in particular for layered superconductors
with weak interlayer coupling such as some high-$T_C$ cuprates. We also 
assume that this coupling does not change the qualitative behavior of our
results obtained for a stacked system of uncoupled layers. For uncoupled 
layers, the basic tenet of dynamical BKT behavior is that vortices and 
antivortices move diffusively with bonding and unbonding of pairs under 
the influence of random internal forces (pinning). We studied the effect of 
the fluctuating magnetic fields accompanying the motion of vortices 
and antivortices on the longitudinal and transverse NQR relaxation rates 
T$_1^{-1}$ and $T_2^{-1}$ in the vicinity of $T_C$. Our procedure uses a
Coulomb gas description of thermally created VA pairs. The NQR relaxation
rates were calculated in terms of the time-dependent correlation functions 
of the fluctuating magnetic field of the vortices and antivortices, 
Eq.~(\ref{CTkaa}). The interaction and recombination of VA pairs were taken 
into account and they can drastically reduce the  correlation time. From the 
results obtained for $T_1^{-1}$ and $T_2^{-1}$, 
Eqs. (\ref{CTrate1}) and (\ref{CTrate2}), it is seen that the vortex diffusion 
constant $D(T)$ plays a crucial role. The reason is that $D(T)$ determines
both the translational motion of the vortices and also the recombination 
time of the VA pairs. From Figs. 9-11 it is seen that $D$ must be sufficiently
small in order to get contributions to the relaxation rates that are 
comparable to those caused by quasi-particle excitations. We discuss 
possible candidates for the experimental NMR spectroscopy of
dynamical BKT effects in Sec. III.  As for the interlayer 
quasi-particle contribution to the relaxation rates, one way to eliminate
these contributions is to observe NQR on interlayer nuclei. Possible
candidates are $^{87}$SR and $^{209}$Bi in Bi-2212 and $^{201}$Hg in
mercury cuprates. Also for NQR experiments on small particles or thin
films, the surface effects can be important because the charge density
oscillations induced by the surface will set up electric field gradients that
can interact with the quadrupolar moments of the nuclei. Hence, an 
unambiguous observation of 2D fluctuation effects in NMR or NQR 
requires thoughtful experimental set-ups.

\acknowledgements

We acknowledge helpful discussions with R. A. Klemm, J. K\"otzler,
and M. Pieper. C.T. acknowledges the support of the Deutsche
Forschungsgemeinschaft.	
\newpage
\newpage
\vspace{0.5in}
  
FIGURE\ CAPTIONS 

1. Temperature regions. The mean field transition temperature of a 2D
layer is given by $T_{C0}$. The regime of long wavelength Gaussian
fluctuations with 2D character lies outside the shaded region of strong
3D fluctuations. Vortex-antivortex fluctuations exist above 
$T_{\text{BKT}}$ in a single layer. In a 3D system of weakly coupled
layers, coming from high temperatures, the 2D fluctuations are cut off 
at the cross-over temperature $T_{cr}$, before the true transition
temperature $T_C$ is reached.

2. Transverse susceptibility diagrams through first order in the pair
fluctuation propagator, $\cal{D}$.

3. The three fluctuation contributions to  
$\mbox{Im}\chi(\omega) = \sum_{\bf q}\mbox{Im}\chi({\bf q},\omega)$
for a clean system without pair breaking measured in units of
$\left[ \left(\frac{a}{\xi_0}\right)^2\frac{\hbar\omega}{E_F}\frac{N_0}{2}\right]$.

4. The fluctuation propagator $\cal{D}$ as a function of momentum 
$x=k/k_F$ for several values of the pair breaking parameter $\alpha_C$.

5. The fluctuation contribution $\mbox{Im}\chi_{FF,\Gamma}^{(1)}(\omega)$ 
measured in units of
$\left[ \left(\frac{a}{\xi_0}\right)^2\frac{\hbar\omega}{E_F}\frac{N_0}{2}\right]$
as a function of  the pair breaking parameter $\alpha_C$ for several 
values of the reduced external frequency 
$\overline{\Omega}=\hbar\omega / 2E_F$. For the experimental value
$\overline{\Omega} = 5 \times 10^{-8}$ the effect of even very little pair 
breaking is drastic.

6. The fluctuation contributions to
$\left[1/(T_1T)\right]_{\mbox{FL}}\,$,
$\mbox{Im}\chi_\Gamma(\omega) =  
\mbox{Im}\chi_{FF,\Gamma}^{(1)} +
 \mbox{Im}\chi_{FF,\Gamma}^{(2)} +  
\mbox{Im}\chi_{GG,\Gamma}^{(1)}$, measured in units of
$\left[ \left(\frac{a}{\xi_0}\right)^2\frac{\hbar\omega}{E_F}\frac{N_0}{2}\right]$
as a function of  the pair breaking parameter $\alpha_C$ for 
$T = 1.03\, T_C$ and 
$\overline{\Omega}=\hbar\omega / 2E_F = 5 \times 10^{-8}$.

7. The fluctuation contribution to
$\left[1/(T_1T)\right]_{\mbox{FL}}\,$, 
$\mbox{Im}\chi_\Gamma(\omega)$, measured in units of
$\left[ \left(\frac{a}{\xi_0}\right)^2\frac{\hbar\omega}{E_F}\frac{N_0}{2}\right]$
as a function of  Temperature for a range of the pair breaking parameter 
$\alpha_C=$ 0.01, 0.02, 0.04, 0.06, 0.08, and 0.10 (from top to bottom)
and $\overline{\Omega}=\hbar\omega / 2E_F = 5 \times 10^{-8}$.
The details for small and large pair breaking are shown in (b) and (c).

8. Energy levels of $^{63}$Cu nuclei in the CuO$_2$ layer due to
quadrupolar splitting. The allowed transitions are indicated by
arrows.

9. The contributions of vortex-antivortex mag\-ne\-tic-field
fluctuations to the longitudinal and transverse relaxation rates
$T_1^{-1}$ and $T_2^{-1}$ as a function of the diffusion constant
$D_{\text{rel}}=2D$. The dotted line gives the contribution from
$k_{xx}(\omega=0)$ to $T_1^{-1}$, corresponding to the first term
on the right hand side of Eq.~(\protect\ref{CTrate1}). The
parameters used in Eqs.~(\protect\ref{CTrate1})--(\protect\ref{CTkaa})
are given in the text.

10. Magnetic-field contribution to the longitudinal relaxation
rate $T_1^{-1}$ near $T_c$ for Bi-2212.

11. Magnetic-field contribution to the transverse relaxation
rate $T_2^{-1}$ near $T_c$ for Bi-2212.

12. Contribution to the longitudinal relaxation rate $T_1^{-1}$
from vortex core relaxation.


\begin{references}
\bibitem{AZdipl2} In Ref.~\onlinecite{AZdipl} the wrong
temperature dependence of $n_p$ is used. The concentration
of free vortex-antivortex pairs $n_p$ does not have a jump at
$T_{\text{BKT}}$.
\bibitem{Aprem0}The diffusion constant $D$ must be sufficiently large
so that $\tau_p$ is small compared with $1/\omega$ for the
magnetic-field fluctuation contribution to be observable; hence
$D\sim 10^{-6}\mbox{ cm$^2$/s}$.
\bibitem{Aprem1}The BKT transition appears to be rather robust against
the weak interlayer coupling in Bi-2212 and against disorder,
cf.~Ref.~\onlinecite{KK}.
\bibitem{Aprem2}The recombination time of a VA pair must be larger
than the typical time scale of the NQR experiment for VA fluctuations
to affect the relaxation rates $T_1^{-1}$ and $T_2^{-1}$. For a similar
case, namely electron spin resonance experiments in semicondutors, the
recombination time for an electron-hole pair ($10^{-4}\mbox{ s}$ in
Si at $300\mbox{ K}$) is much larger than the time scale of the ESR
experiment ($10^{-12}$ to $10^{-11}\mbox{ s}$).
\end{references}
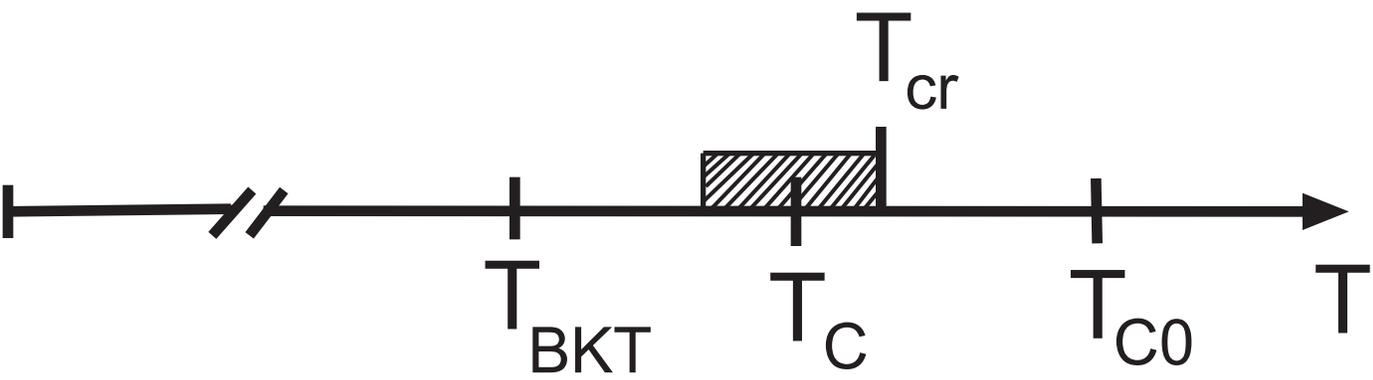
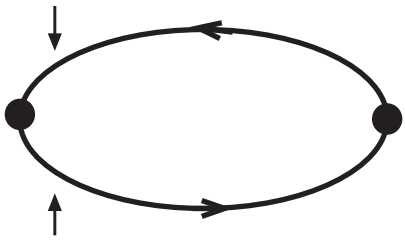
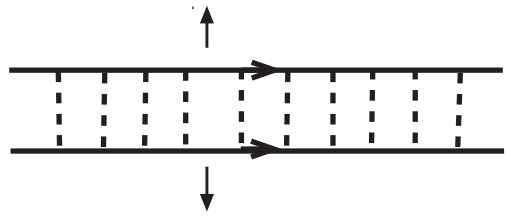
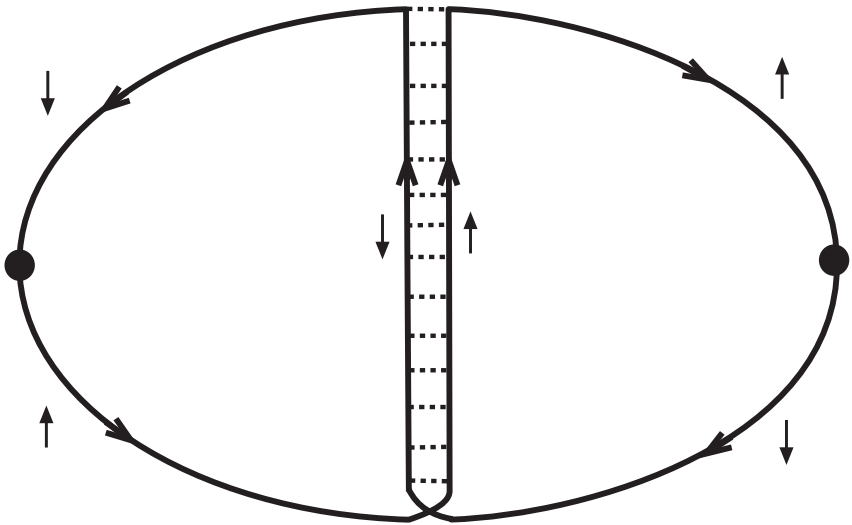
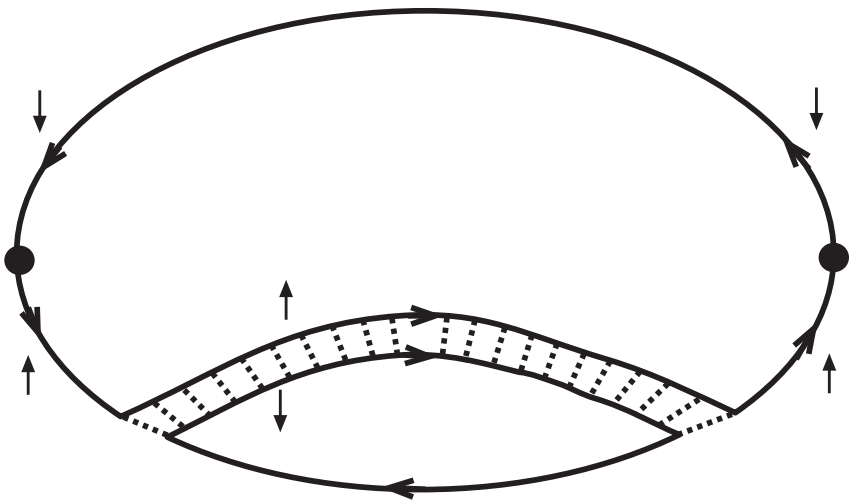
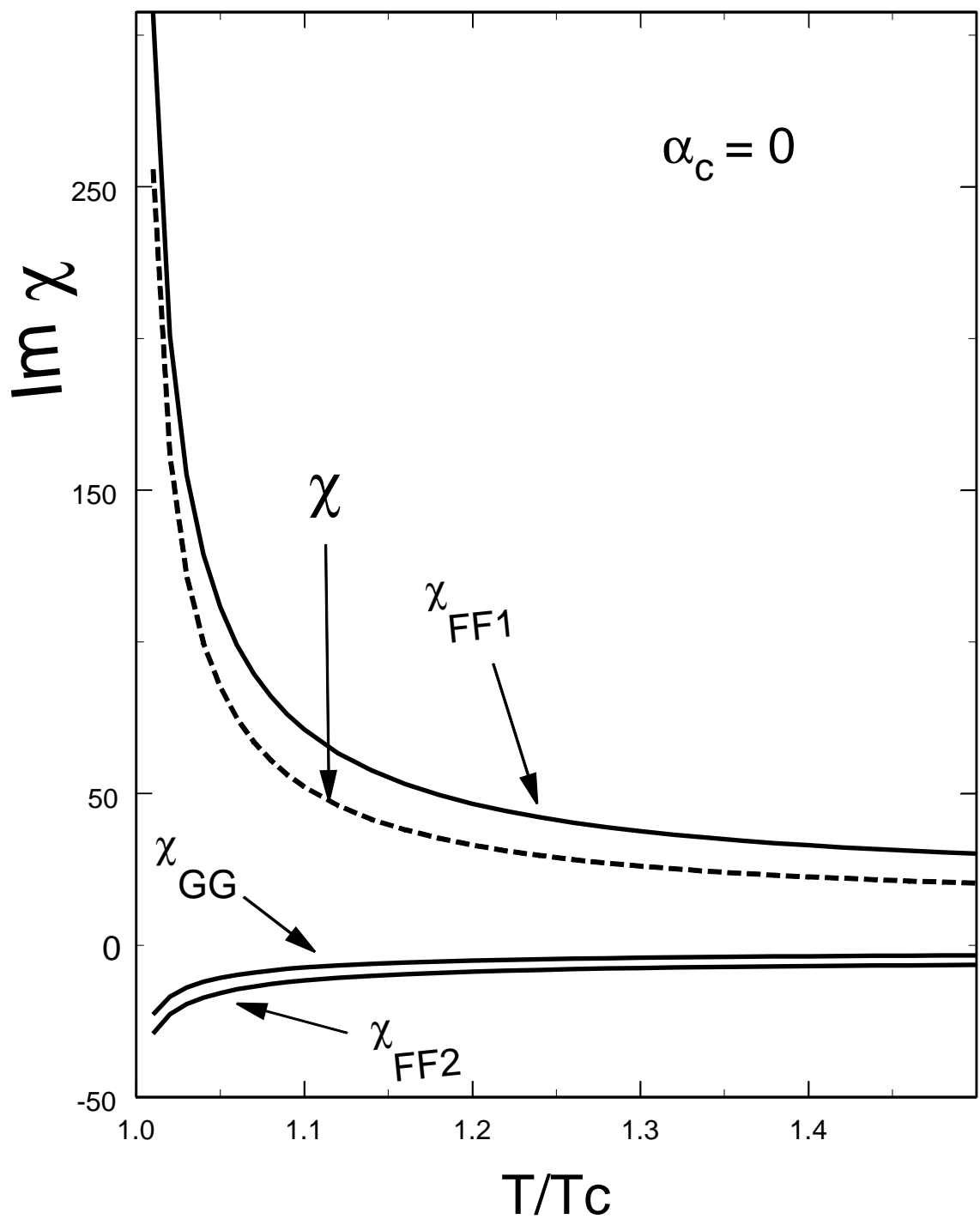
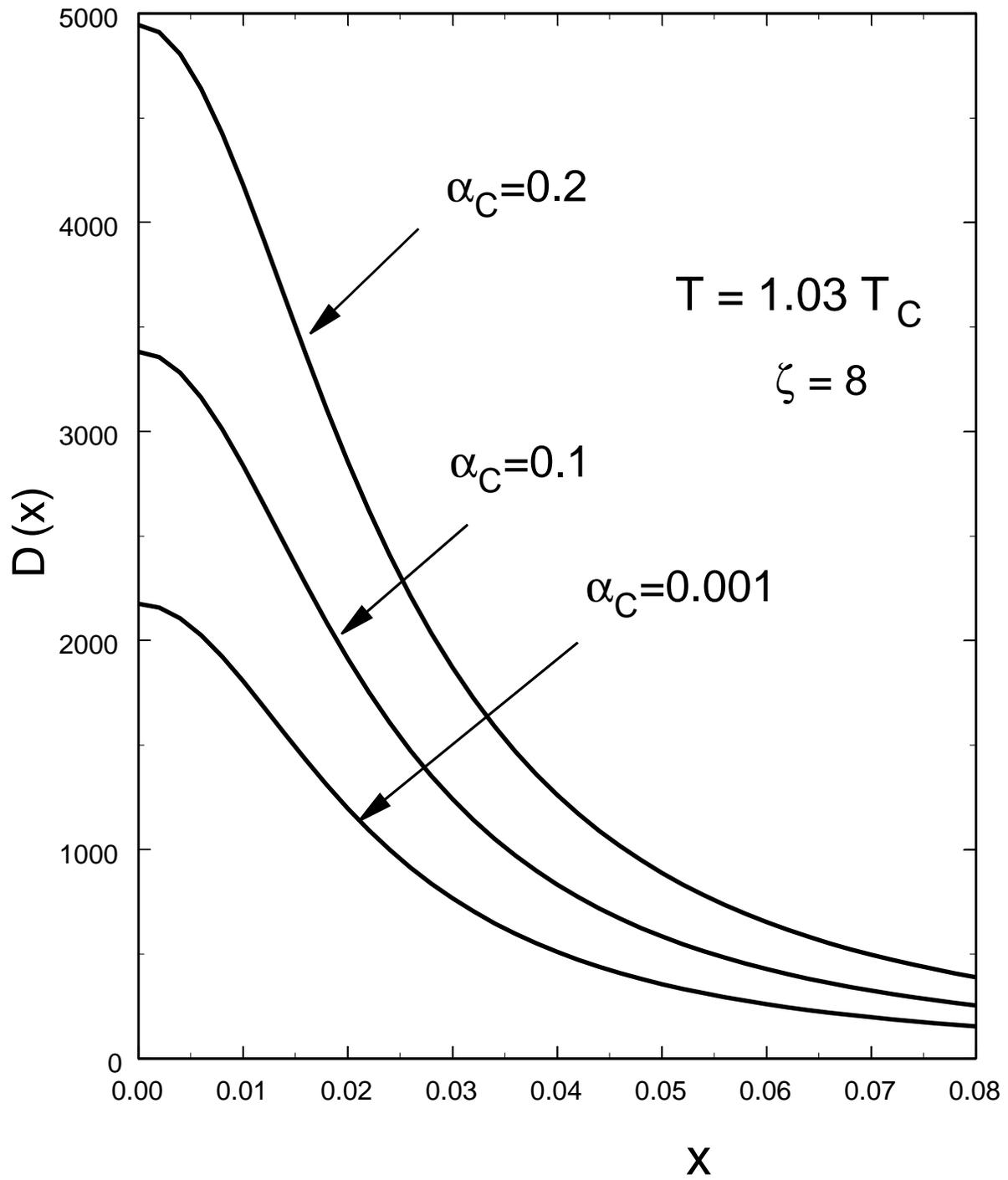
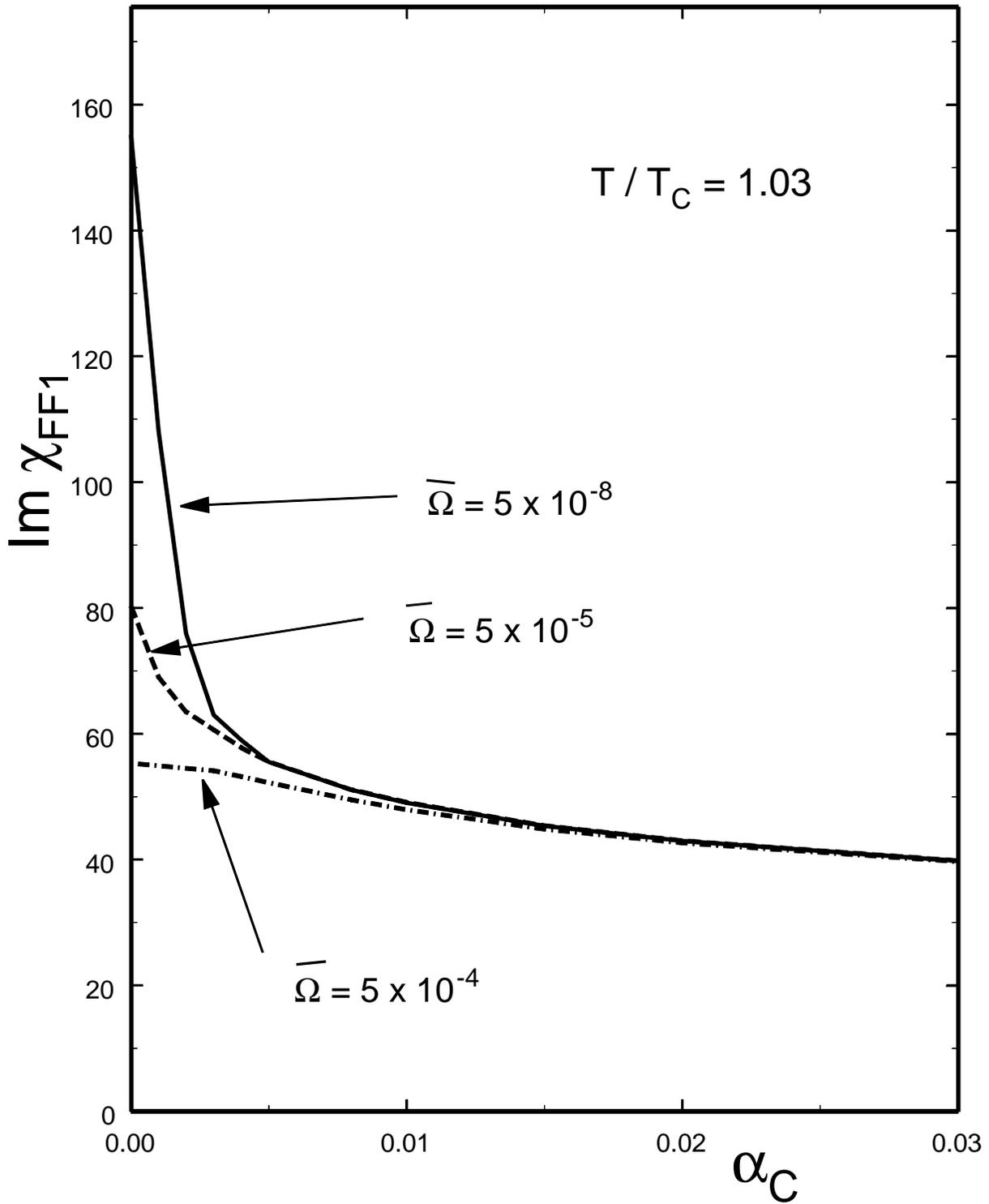
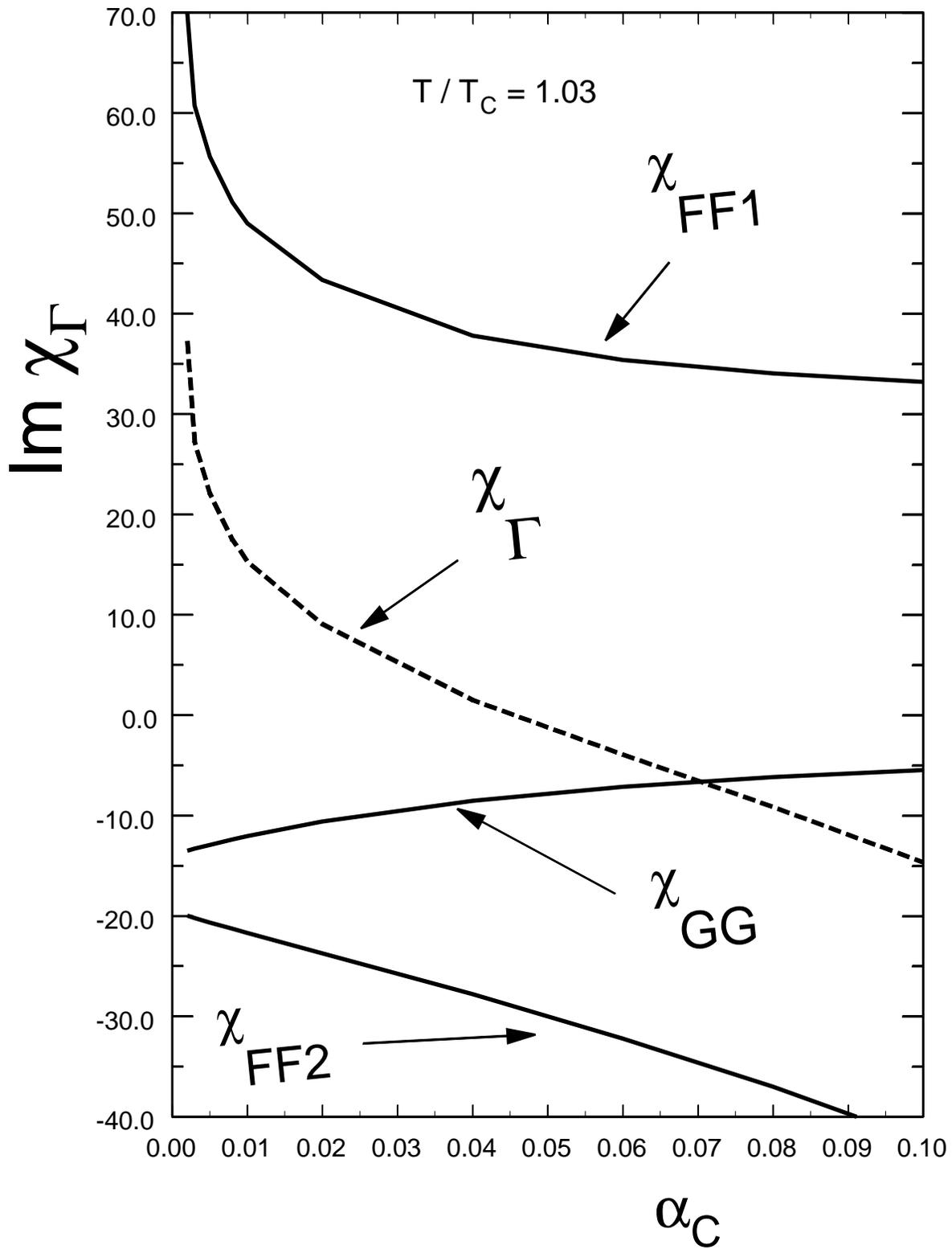
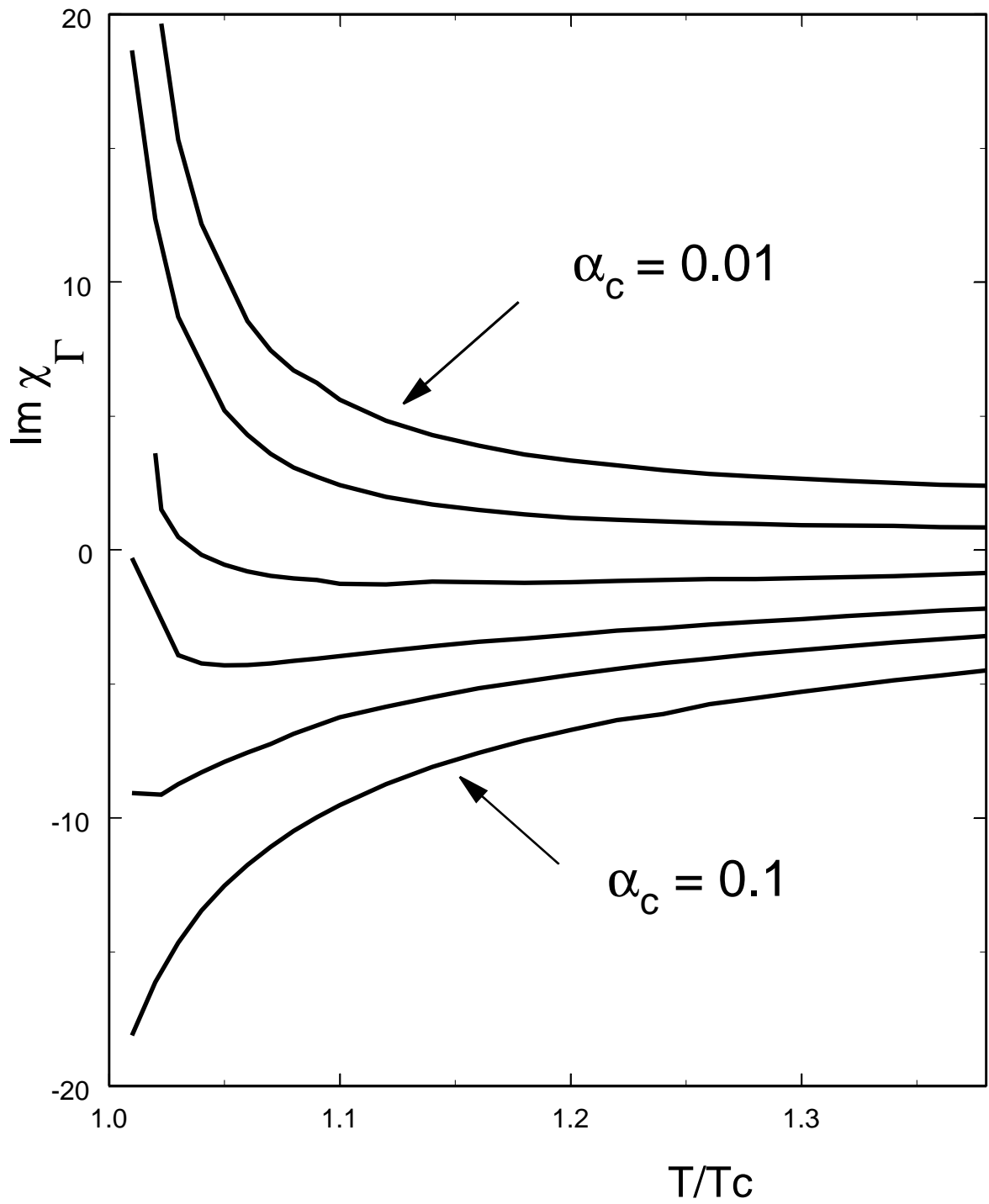
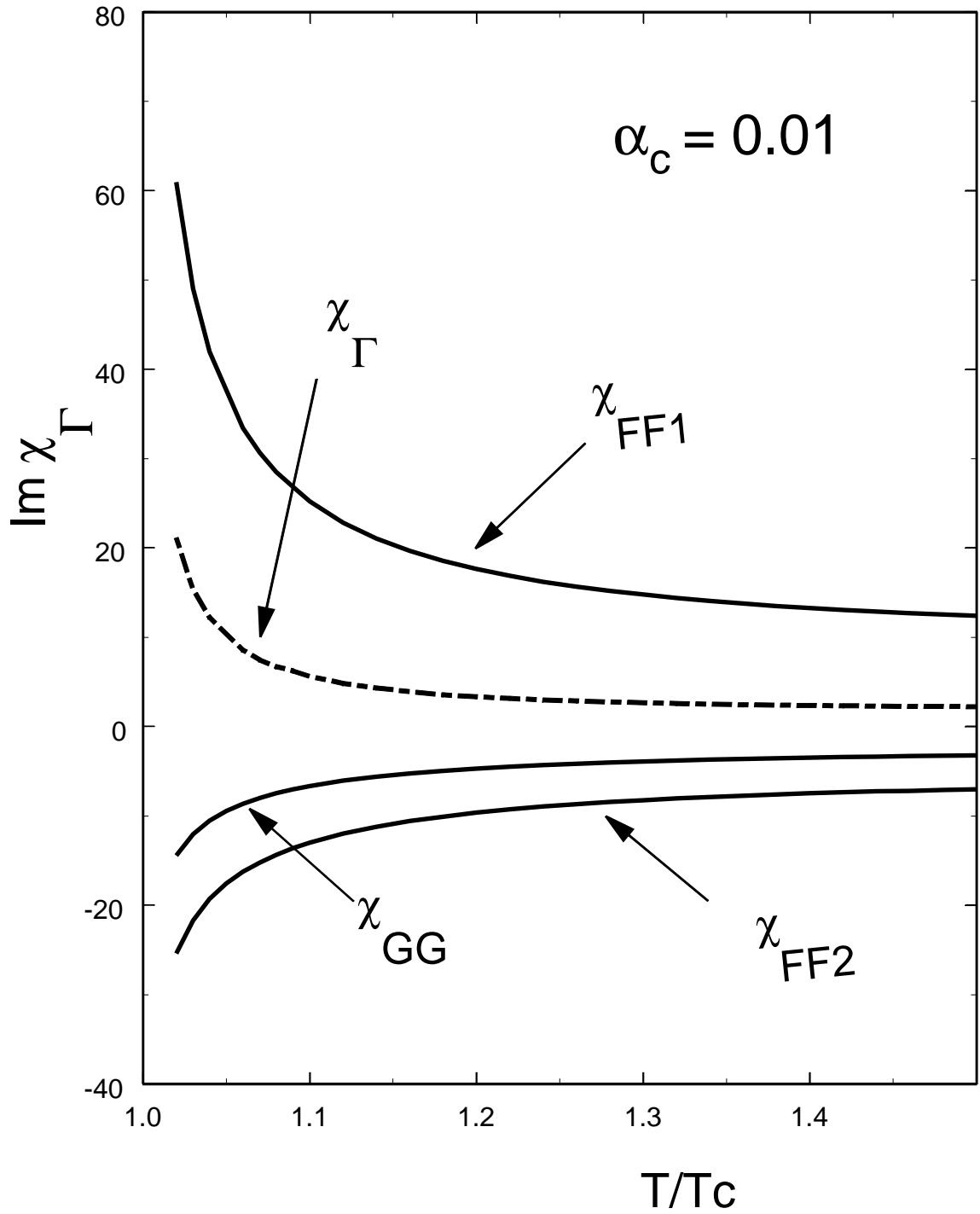
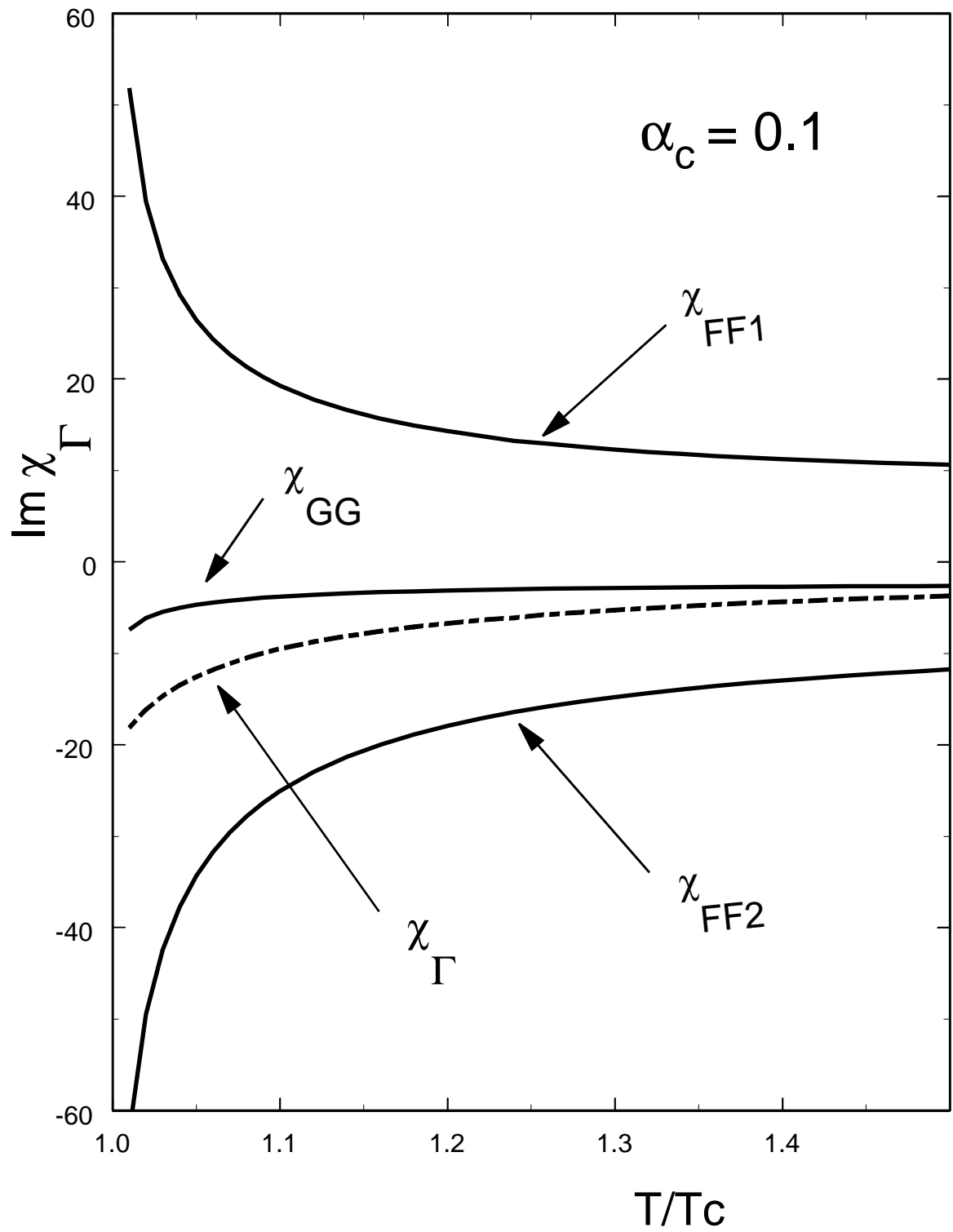
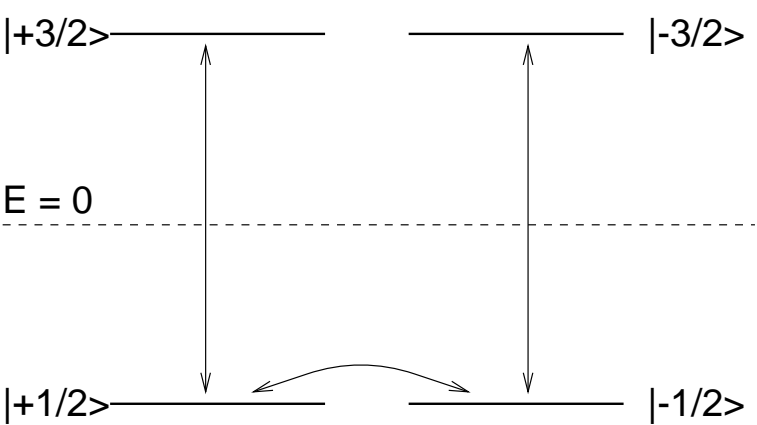
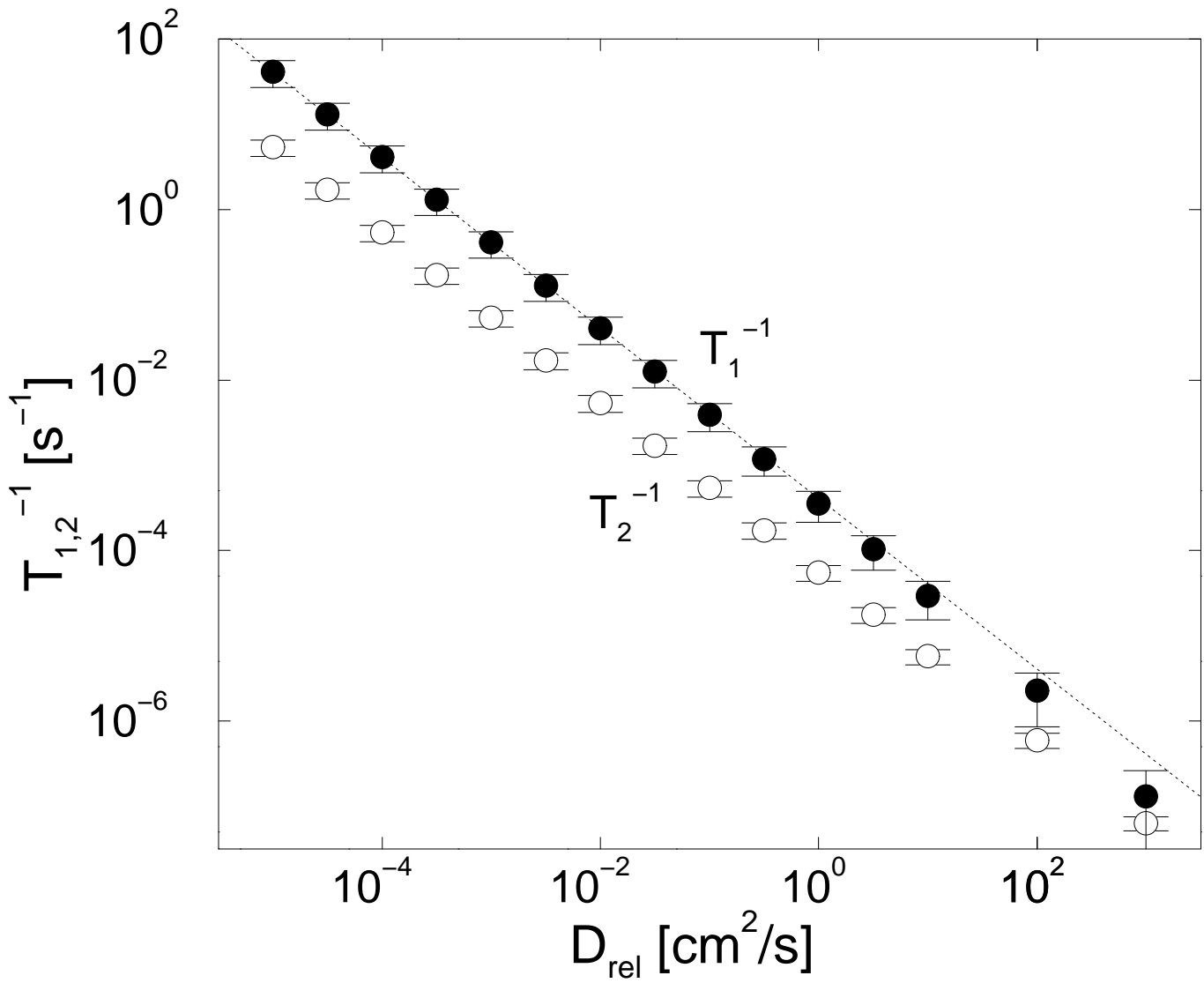
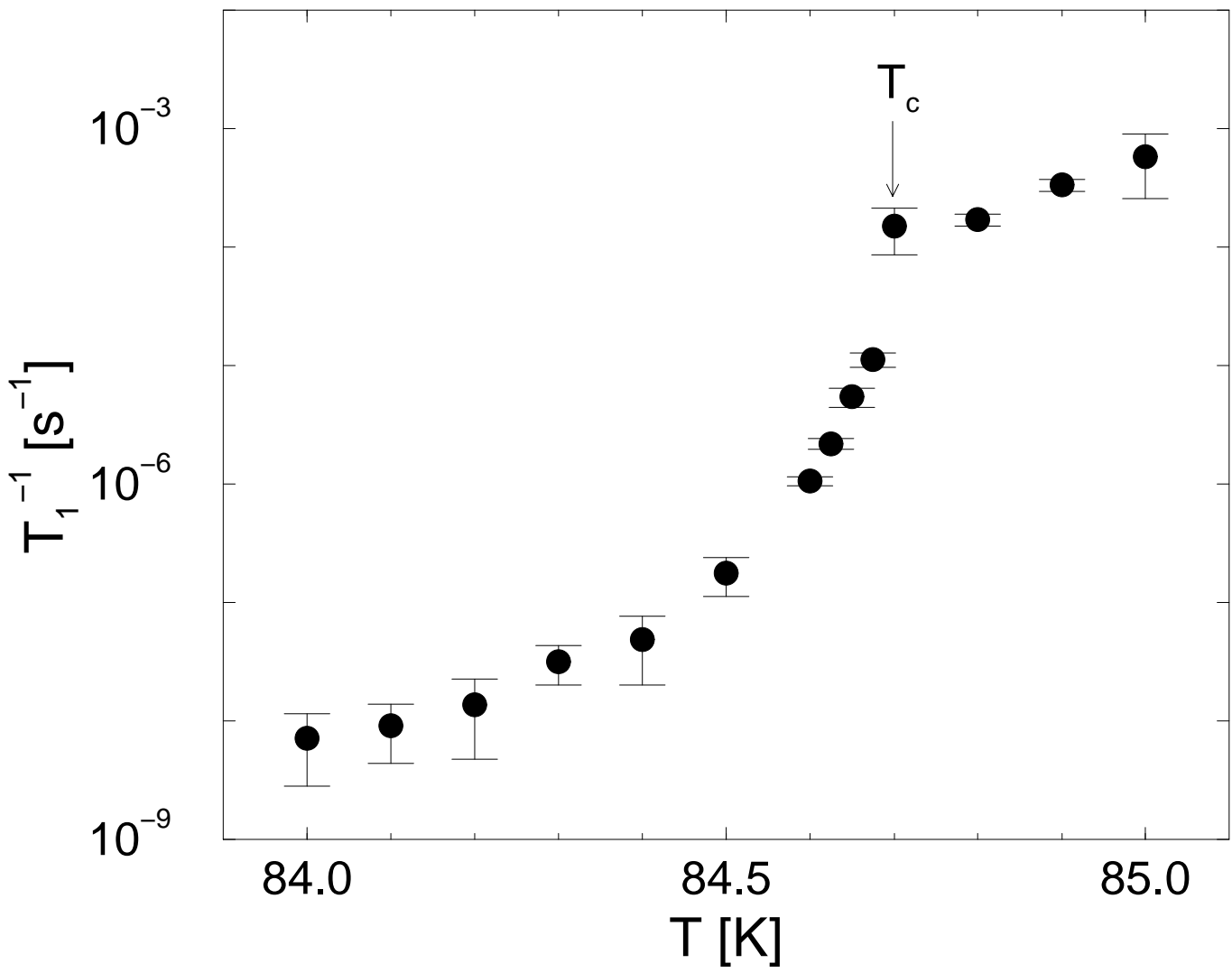
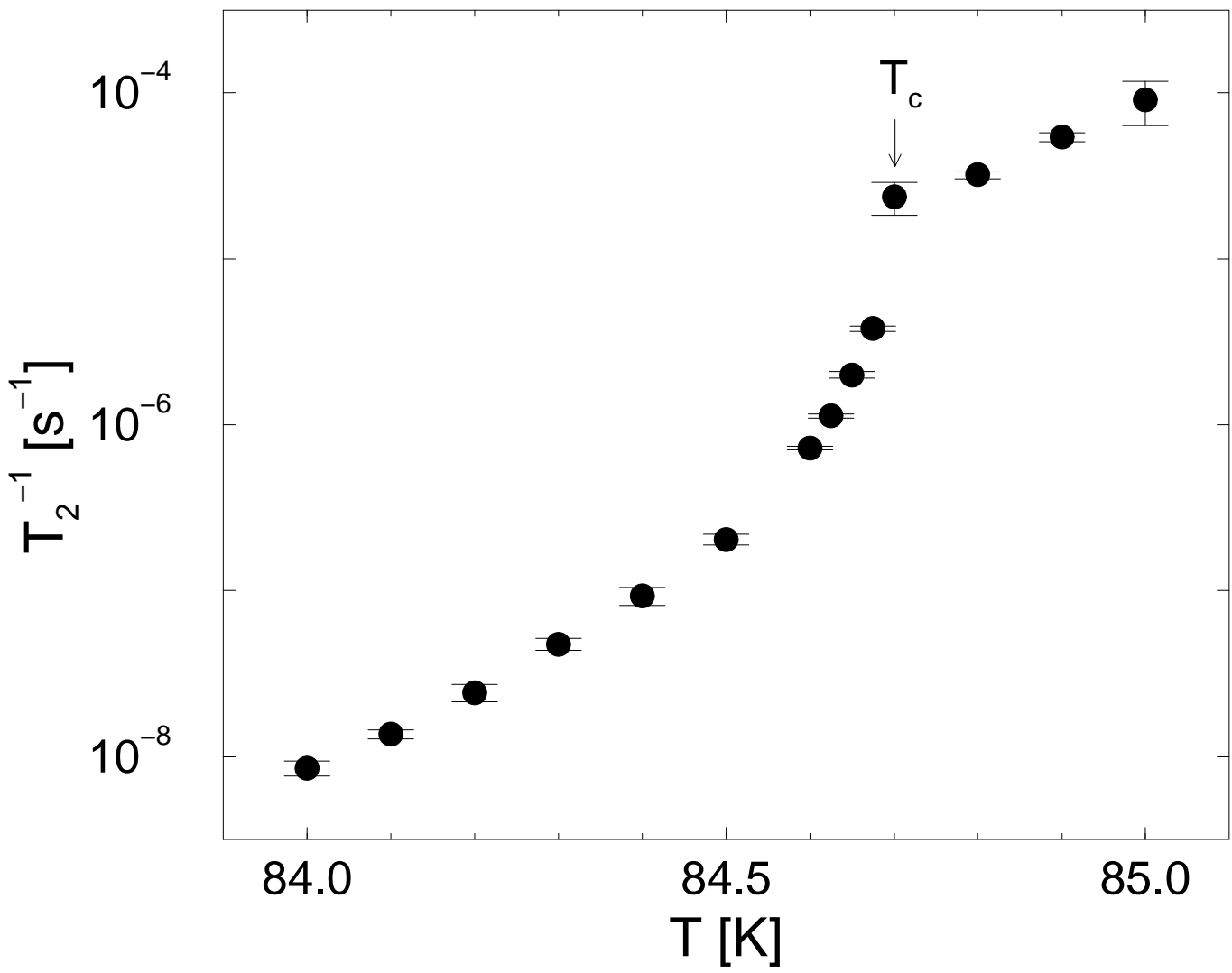
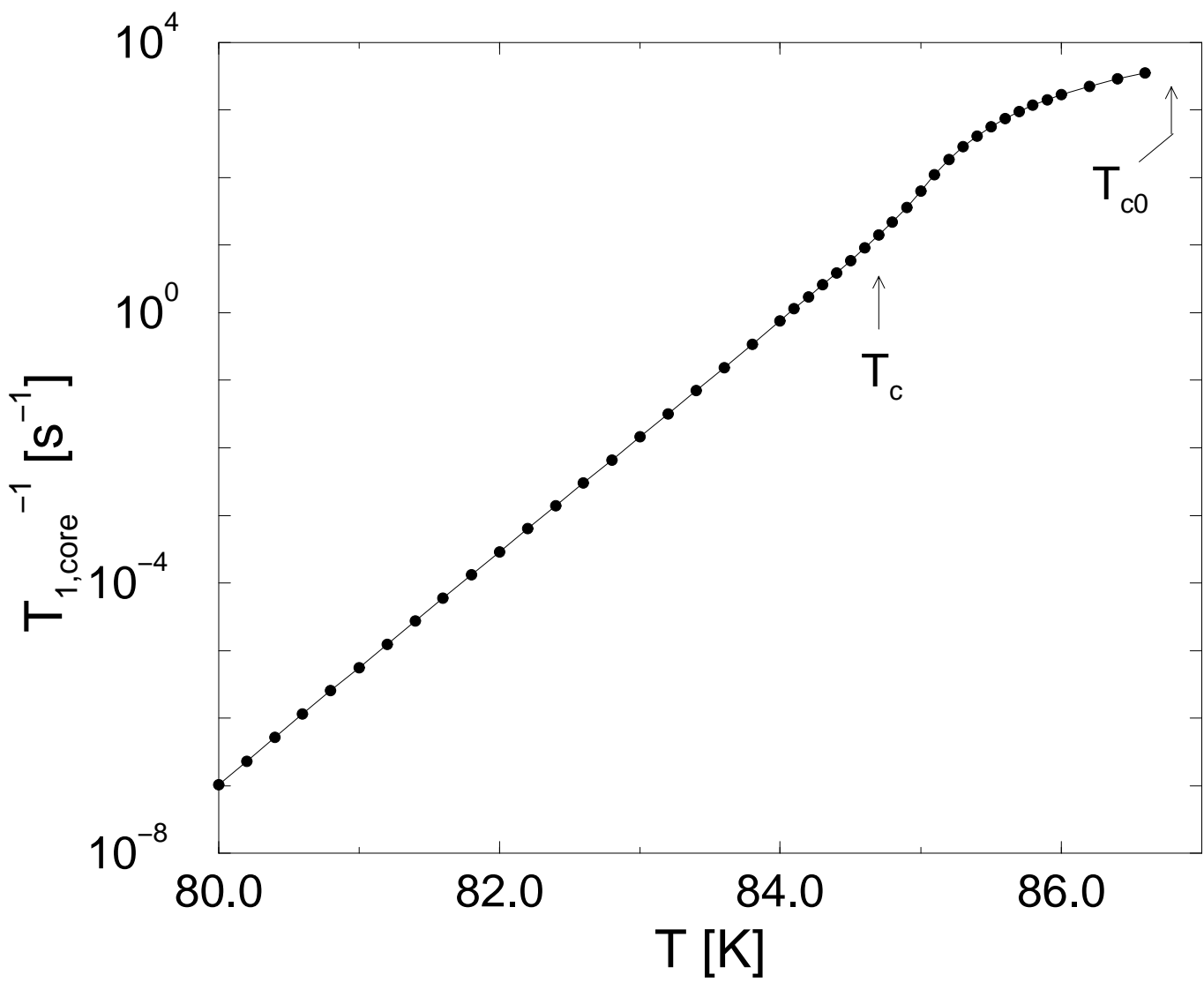
\end{document}